\documentclass[12pt]{elsarticle}

\pdfoutput=1

\usepackage{lipsum}
\makeatletter
\def\ps@pprintTitle{%
	\let\@oddhead\@empty
	\let\@evenhead\@empty
	\def\@oddfoot{}%
	\let\@evenfoot\@oddfoot}
\makeatother

\usepackage{amssymb}
\usepackage{bm}
\usepackage{amsmath}
\usepackage{amsthm}
\usepackage{algorithm}
\usepackage{algorithmic}
\usepackage{booktabs}
\usepackage{hyperref}
\usepackage[normalem]{ulem}
\usepackage{soul}
\usepackage{subfigure}
\usepackage{graphicx}
\usepackage{caption}
\usepackage{xcolor}
\usepackage{multirow}
\usepackage[draft,inline,nomargin,index]{fixme}
\fxsetup{theme=color,mode=multiuser}

\definecolor{mygreen1}{rgb}{0.2,0.6,0.2}
\definecolor{myblue}{rgb}{0,0.38,0.52}
\FXRegisterAuthor{hl}{ahl}{\color{myblue}Han}
\FXRegisterAuthor{mk}{amk}{\color{mygreen1}Moe} 
\FXRegisterAuthor{cs}{acs}{\color{blue}Cosmin} 
\newcommand{\cs}[1]{\color{purple}\,Cosmin:{#1}\color{black}}

\newtheorem{lemma}{Lemma}

\journal{Computer Methods in Applied Mechanics and Engineering}

\begin{document}

\begin{frontmatter}



\title{Generalized Transitional Markov Chain Monte Carlo Sampling Technique for Bayesian Inversion}


\address[label1]{University of Houston, Houston, Texas 77204, USA}
\address[label2]{Sandia National Laboratories, Livermore, California 94550, USA}
\address[label3]{Cyentech Consulting LLC, Cypress, Texas 77429, USA}

\author[label1]{Han Lu}

\author[label2]{Mohammad Khalil}
\ead{mkhalil@sandia.gov}

\author[label2]{Thomas Catanach}

\author[label1]{Jiefu Chen}

\author[label1]{Xuqing Wu}

\author[label1]{Xin Fu}

\author[label2]{Cosmin Safta}

\author[label3]{Yueqin Huang}

\begin{abstract}
In the context of Bayesian inversion for scientific and engineering modeling, Markov chain Monte Carlo sampling strategies are the benchmark due to their flexibility and robustness in dealing with arbitrary posterior probability density functions (PDFs). However, these algorithms been shown to be inefficient when sampling from posterior distributions that are high-dimensional or exhibit multi-modality and/or strong parameter correlations. In such contexts, the sequential Monte Carlo technique of transitional Markov chain Monte Carlo (TMCMC) provides a more efficient alternative. Despite the recent applicability for Bayesian updating and model selection across a variety of disciplines, TMCMC may require a prohibitive number of tempering stages when the prior PDF is significantly different from the target posterior. Furthermore, the need to start with an initial set of samples from the prior distribution may present a challenge when dealing with implicit priors, e.g. based on feasible regions. Finally, TMCMC can not be used for inverse problems with improper prior PDFs that represent lack of prior knowledge on all or a subset of parameters. In this investigation, a generalization of TMCMC that alleviates such challenges and limitations is proposed, resulting in a tempering sampling strategy of enhanced robustness and computational efficiency. Convergence analysis of the proposed sequential Monte Carlo algorithm is presented, proving that the distance between the intermediate distributions and the target posterior distribution monotonically decreases as the algorithm proceeds. The enhanced efficiency associated with the proposed generalization is highlighted through a series of test inverse problems and an engineering application in the oil and gas industry.
\end{abstract}


\end{frontmatter}



\section{Introduction}
\label{sec:intro}

Bayesian inference, an application of Bayes' theorem for statistical inference, has been increasingly applied in many scientific and engineering disciplines. In such settings, Bayesian inference provides the means to update predictive models of the physical process of interest using available observations while accounting for uncertainties in the observations, model parameters and model structure. Among various Bayesian inference methods, Markov chain Monte Carlo (MCMC) sampling~\cite[]{gilks1995markov, brooks2011handbook} is the most popular method due to its reliability, generality, and ease of implementation~\cite[]{dunn2011exploring}. However, traditional MCMC methods, such as Metropolis-Hastings (MH)~\cite[]{hastings1970monte}, can be inefficient when the target posterior probability density function (PDF) is (a) high dimensional~\cite[]{au2003important, cotter2013mcmc, roberts2001optimal}, (b) exhibits multi-modality~\cite{robert2018accelerating}, or (c) when the unknown variables are highly correlated~\cite[]{beck2002bayesian} such that the posterior PDF approximates a low-dimensional manifold in parameter space. To alleviate such limitations, a variety of MCMC variants have been developed such as adaptive MCMC~\cite{roberts2009examples,haario2006dram}, parallel tempered MCMC~\cite{desjardins2010parallel}, and hybrid Monte Carlo~\cite{duane1987hybrid}. In particular, the transitional MCMC (TMCMC) algorithm was first proposed for applications of Bayesian inference in structural dynamics and other engineering applications~\cite{ching2007transitional}. TMCMC, and associated extensions, are generally classified as Sequential Monte Carlo (SMC) methods which evolve populations of samples through a sequence of distributions \cite{del2006sequential, kantas2014sequential, beskos2016convergence,marinari1992simulated}. Instead of directly sampling from the target PDF, TMCMC generates samples from a series of intermediate PDFs that gradually evolve from the prior to the posterior distribution, and are focused on efficiently concentrating the samples in regions of parameter space with high probability mass according to the posterior distribution. At each iteration of TMCMC, the likelihood function can be independently evaluated across all samples, leading to naturally parallelizable sampling strategies as opposed to standard MCMC methods that evaluate the likelihood sequentially. The ability to sample from complex target PDFs in parallel has promoted TMCMC, and its modified versions, for use in estimating model parameters in many areas of application~\cite{ortiz2015identification, betz2016transitional, ching2016application, lee2017system, wu2018bayesian,jin2019identifying, catanach2018bayesian, catanach2020bayesian, Ching:2021}.

Parallel tempering (PT)~\cite{swendsen1986replica,geyer1991markov,marinari1992simulated,earl2005parallel} is another approach that is similar to TMCMC in two aspects: (a) deal with a set of intermediate PDFs with associated tempering parameter and (b) provide means of achieving parallelism through multiple MCMC chains. However, the main difference between the two sets of algorithms relate to the means of achieving parallelism: TMCMC runs multiple MCMC chains for each stage (with an associated tempering parameter) whereas PT runs multiple chains, one for each tempering parameter value. In practice, one must start with a pre-specified temperature schedule for PT (in an offline phase) whereas TMCMC learns the optimal temperature schedule based on the target PDF of interest. Furthermore, unlike many MCMC algorithms, TMCMC does not normally suffer from burn-in, the generation of samples at the initial stages of an MCMC algorithm that are not distributed according to the target or stationary PDF. The samples that are generated during the burn-in phase are typically discarded, reducing the overall efficiency of MCMC. TMCMC does not normally have a burn-in period since the initial distribution is the prior PDF as opposed to an arbitrary distribution associated with initial sample generation for MCMC. Another important feature of TMCMC is that it provides an estimate of the Bayesian model evidence~\cite{sivia2006data}, an essential quantity in Bayesian model selection \cite{bisaillon2015bayesian,sandhu2017bayesian} and model averaging \cite{hoeting1999bayesian,wasserman2000bayesian}, at normally negligible additional cost.

In applications of TMCMC to Bayesian inference, the initial samples are drawn from the prior PDF which characterizes the initial knowledge, or lack thereof, of the parameter vector. Those samples are transitioned from one stage/intermediate PDF to the next using a combination of a sampling importance resampling \cite{asmussen2007stochastic} step and MCMC simulations, finally arriving at an ensemble that is distributed according to the posterior PDF. TMCMC has been shown to be effective in dealing with posterior PDFs that are high-dimensional, exhibit multi-modality and/or strong parameter correlations. This has prompted recent efforts towards the development of algorithms that are modifications of or related to TMCMC, such as algorithms that deal with ``big data" \cite{green2015bayesian,green2015bayesianb} and algorithms that couple TMCMC with more efficient MCMC samplers \cite{angelikopoulos2015x,jin2019identifying}. Nonetheless, existing TMCMC algorithms still face computational and formulation issues that might limit their applicability in practical applications of Bayesian inference. 

For a pre-specified sampling error budget, a prior PDF that is somewhat ``closer" to the posterior PDF often reduces the number of TMCMC tempering stages and the associated forward model simulations for likelihood evaluation. The opposite is also true in that a prior PDF that is too diffuse with respect to the posterior would place prohibitive computational costs due to increasing number of stages to achieve the same level of sampling error. In Bayesian calibration tasks, modelers might assign improper priors to parameters to represent the lack of knowledge. \textit{TMCMC can not be applied to Bayesian calibration tasks involving such uninformative improper prior PDFs due to the inability to provide an initial set of samples to TMCMC}. There are also scenarios in which the prior knowledge on unknown parameters is in the form of constraints that are strongly tied to the physical laws associated with the system being modeled. In such settings, the prior PDF takes the form of a uniform density on the feasible region being the set of all possible points in parameter space that satisfy such constraints. In such scenarios, direct Monte Carlo sampling of such ``implicit" prior PDFs becomes challenging, especially as the number of constraints, the degree of constraint non-linearities, and/or the dimensionality (i.e. number of unknown parameters) increase. Although sampling from a feasible set may be achieved using importance sampling (IS), sampling importance resampling as well as rejection sampling \cite{gelman2013bayesian}, there might be situations when the feasible set is unbounded, leading again to improper prior PDFs. The modeler might feel compelled to further constraint the parameters to arrive at a proper prior PDF in order to use TMCMC, but such constraints might strongly influence the posterior PDFs and subsequent model predictions. 

The SMC approach presented herein is a generalization of TMCMC that allows the algorithm to start with an arbitrary initial state, an importance PDF, instead of the traditional prior PDF as is the case for classical variants of TMCMC. The proposed generalized TMCMC (GTMCMC) technique is a tempering algorithm~\cite{marinari1992simulated} in that it generally allows both cooling and heating of the solution depending on the choice of importance PDF with respect to the target posterior PDF. The number of TMCMC stages, and associated forward model simulations for likelihood evaluation, may be dramatically reduced when the importance distribution is chosen judiciously to more closely resemble the target posterior PDF than does the prior PDF. Furthermore, the importance distribution may be chosen to be one that is easy to generate samples from in comparison to prior PDFs that might be based on feasible sets. The ability to start with samples from an arbitrary importance PDF would also extend the applicability of the proposed TMCMC algorithm to situations involving improper prior PDFs, a major limitation of traditional TMCMC algorithms. It is also important to note that when the importance PDF is chosen as the prior PDF, the proposed algorithm behaves exactly as the traditional TMCMC algorithm. Therefore, this approach can be considered as a generalization of the original TMCMC sampling technique. The proposed GTMCMC algorithm addresses several challenges/inefficiencies that standard TMCMC suffers from towards providing a more robust and efficient tempering sampling strategy, namely:

\begin{enumerate}
	\item Bayesian inference problems involving improper priors which prevent the use of TMCMC
	\item Difficult to sample priors, including implicit priors, where the prior is not explicitly defined as a function of parameters but as a function of some other quantifies of interests and therefore can typically only be sampled through rejection sampling and other similar approaches.
    \item Solving sequences of related Bayesian inference problems where parameter posterior PDFs are "close" to previously obtained solutions of similar inverse problems (in contrast to prior PDFs). This concept is similar to training tasks in machine learning whereby transfer learning~\cite{baxter1997bayesian} is used.
\end{enumerate}

The remainder of this paper is organized as follows. Section~\ref{sec:BayesianInference} serves as an introduction into Bayesian inference for scientific and engineering system. Section~\ref{sec:TMCMC} overviews TMCMC sampling for Bayesian inversion and Section~\ref{sec:GTMCMC} presents the proposed GTMCMC sampling algorithm. In Section~\ref{sec:NumTests}, we demonstrate the advantages of the new algorithm over the standard TMCMC through several comparative numerical investigations. In Section~\ref{sec:LWD}, we apply the proposed algorithm on an engineering application in the oil and gas industry: the inversion of the ultra-deep directional resistivity data obtained from the logging-while-drilling tool. Finally, we provide concluding remarks and observations regarding the proposed algorithm in Section~\ref{sec:conclusion}.

\section{Bayesian Inference}
\label{sec:BayesianInference}
Given a forward model $f$, unknown parameter vector $\bm{\theta} \in \mathbb{R}^{N_\theta}$, and experimental parameter vector $\bm{x}$, the noisy observation vector ${\bm{d}_j}\in \mathbb{R}^{N_d}$ may be given by 
\begin{equation} \label{forward_model}
    {\bm{d}_j} = f(\bm{\theta}, \bm{x}) + \epsilon_j \ ,
\end{equation}
in which $\bm{\epsilon_j}$ is a specific realization of the measurement error vector $\epsilon\in \mathbb{R}^{N_\epsilon}$. Given a set of noisy measurements $\bm{D} = \left\{ \bm{d}_1, \hdots, \bm{d}_N \right\}$, associated with pre-specified experimental parameter vector realizations $\bm{X} = \left\{ \bm{x}_1, \hdots, \bm{x}_N \right\}$ the inverse problem associated with inferring the unknown (or weakly known) parameter vector $\bm{\theta}$ may be solved using Bayesian inference~\cite{stuart2010inverse}, with Bayes' law providing the solution in the form of a joint PDF given by
\begin{equation}\label{posterior}
	p(\bm{\theta} \vert \bm{D}) \propto p(\bm{D} \vert\bm{\theta})p(\bm{\theta}) \ ,
\end{equation}
where $p(\bm{\theta} \vert \bm{D}) $ represents posterior PDF, $p(\bm{D}\vert\bm{\theta})$ is the likelihood of the data $\bm{D}$ for a given realization of the parameter vector $\bm{\theta}$, and $p(\bm{\theta})$ is the prior PDF that encapsulates any knowledge about the parameter vector prior to assimilating the available data.

Under specific circumstances in which the forward model is linear in both the unknown parameter vector $\theta$ and the measurement error vector, and with a multivariate Gaussian measurement error vector and Gaussian prior PDF, the posterior PDF is exactly Gaussian. Generally, the posterior is not Gaussian and the analytical solution of the posterior distribution is intractable. In such situations, a popular approach is to apply MCMC sampling to draws a series of samples from $p(\bm{\theta} \vert {\bm{D}})$ by constructing a Markov chain that has the target posterior as its equilibrium/stationary distribution~\cite{roberts2009examples,haario2006dram}.

In practical scenarios, many samples ($1 \times 10^6$ or more) are needed to effectively explore the parameter space in regions associated with high probabilities for subsequent predictive tasks. This issue is exacerbated by the fact that MCMC methods produce correlated, rather than independent, samples, with an associated correlation length that tends to increase with increasing problem dimensionality. Many efforts have focused designing more effective proposal distributions that aim to decrease the correlation length associated with the samples. However, to date, optimal proposal distributions are only available in simplistic settings~\cite{roberts2001optimal,rosenthal2011optimal}. Therefore, it is necessary to explore more efficient sampling methods for such problems. Furthermore, MCMC relies on sequential evaluation of the posterior PDF and associated likelihood function through forward model simulations due to the reliance on a Markov chain for sampling. This limits the applicability of MCMC towards solving inverse problems involving computationally intensive forward model simulations as it limits the use of high performance computing to evaluate samples in parallel.

\section{Transitional Markov chain Monte Carlo}
\label{sec:TMCMC}

While MCMC draws samples from the target PDF in a sequential manner, TMCMC generates samples from a series of intermediate PDFs that gradually evolve from the prior to the posterior (target) distribution, and are focused on efficiently concentrating the samples in regions of parameter space with high probability mass according to the posterior distribution. At each iteration of TMCMC, the likelihood function can be independently evaluated across all samples, leading to naturally parallel sampling strategies as opposed to sequential MCMC methods. Various investigations have demonstrated the enhanced efficiency of TMCMC, in comparison to MCMC, when generating samples from complex target posterior PDFs~\cite{ortiz2015identification, betz2016transitional, ching2016application, lee2017system, wu2018bayesian,jin2019identifying, catanach2018bayesian, catanach2020bayesian}. Additionally, TMCMC provides an estimate for the Bayesian model evidence~\cite{sivia2006data} with minimal additional cost without further model (likelihood) evaluations.

As previously mentioned, TMCMC constructs and samples from a series of intermediate PDFs. Consider the following intermediate PDFs
\begin{equation}
    p_j(\bm{\theta}) \propto p(\bm{D}\vert\bm{\theta})^{\beta_j}p(\bm{\theta}) \ ,
\end{equation}
with $j = 0,\hdots,m$ and $0=\beta_0<\beta_1<\hdots<\beta_m=1$, where index $j$ denotes the TMCMC stage number associated with a specific tempering parameter $\beta_j$. The intermediate PDFs $p_j$ transition from the prior PDF $p(\bm{\theta})$ towards the posterior PDF $p(\bm{\theta}\vert \bm{D})$. Though the topology change from $p(\bm{\theta})$ to $p(\bm{\theta}\vert \bm{D})$ can be dramatic, the change between two subsequent intermediate PDFs can be gradual through a judicious selection of $\beta_j$, thus making it possible to efficiently "nudge" or transition the samples from one stage to the next. We summarize the TMCMC algorithm in Algorithm~\ref{tmcmc}.
\begin{algorithm}[H]
\footnotesize
\begin{algorithmic}
\caption{Transitional Markov chain Monte Carlo} 
\label{tmcmc}
\REQUIRE Prior PDF, $p(\bm{\theta})$; Likelihood function, $p(\bm{D} \vert \bm{\theta})$; Number of samples, $n$; Target coefficient of variation, $CoV$, of the weights; Prescribed scaling factor for MCMC proposal covariance matrix, $\gamma$; MCMC chain length, $N$ 
\ENSURE Samples from posterior PDF, $\{\bm{\theta}_1, \hdots, \bm{\theta}_n \}$ \& Model evidence estimate, $S$ 
    \STATE $\beta_0 \leftarrow 0$\; $k \leftarrow 0$\;
    \STATE Generate samples $\{\bm{\theta}_1^{(0)}, \hdots, \bm{\theta}_n^{(0)} \}$ from $p(\bm{\theta})$\;
    \WHILE{$\beta_k < 1$}
        	\STATE Compute likelihood of ensemble of realizations $p(\bm{D} \vert \bm{\theta}_l^{(k)})$, $l = 1, \hdots, n$\;
        	\STATE Compute the weight of each sample $w(\bm{\theta}_l^{(k)})=p(\bm{D}\vert \bm{\theta}_l^{(k)})^{\beta_{k+1} - \beta_{k}}$, with $\beta_{k+1}$ chosen such that  coefficient of variation of those weights is equal to the target $CoV$\;
        	\STATE Compute contribution to model evidence, $S_k = \frac{1}{n}\sum_{l=1}^{n}w(\bm{\theta}_l^{(k)})$\;
        	\STATE Compute the ensemble covariance matrix to be used in MCMC simulations:
        	\begin{align*}
        	\tilde{w}(\bm{\theta}_l^{(k)}) = \frac{w(\bm{\theta}_l^{(k)})}{\sum_{l=1}^{n}w(\bm{\theta}_l^{(k)})} \ \ \ \ \ \ \ \overline{\bm{\theta}}^{(k)} = \sum_{k=1}^{n}\bm{\theta}_l^{(k)} \tilde{w}(\bm{\theta}_l^{(k)})\\
        	\bm{\Sigma}^{(k)} = \gamma^2\sum_{k=1}^{n}\tilde{w}(\bm{\theta}_l^{(k)})(\bm{\theta}_l^{(k)} - \overline{\bm{\theta}}^{(k)})(\bm{\theta}_l^{(k)} - \overline{\bm{\theta}}^{(k)})^T
        	\end{align*}\\
        	\STATE Resampling: Generate $n$ random samples, $\{\tilde{\bm{\theta}}_1^{(k)}, \hdots, \tilde{\bm{\theta}}_n^{(k)} \}$, from the current ensemble $\{\bm{\theta}_1^{(k)}, \hdots, \bm{\theta}_n^{(k)} \}$ with corresponding probabilities $\tilde{w}(\bm{\theta}_l^{(k)})$\;
        	\FOR{$l\leftarrow 1$ \textbf{to} $n$}
        		\STATE $\bm{x}_0 \leftarrow \tilde{\bm{\theta}}_l^{(k)}$\;
        		\FOR{$i\leftarrow 0$ \textbf{to} $N-1$}
	        		\STATE Generate $\hat{\bm{x}}$ from Gaussian proposal $\mathcal{N}(\bm{x}_i, \bm{\Sigma}^{(k)})$\;
	        		\STATE Compute acceptance probability, $\alpha = {\rm min} \{1, \frac{p(\hat{\bm{x}}) p(\bm{D} \vert \hat{\bm{x}})^{\beta_{k+1}}}{p(\bm{x}_i) p(\bm{D} \vert \bm{x}_i)^{\beta_{k+1}}}\}$\;
	        		\STATE Generate $u$ from from the continuous uniform distribution on the interval  $\left[0,1\right]$\;
	        		\IF{$u < \alpha$}
	        			\STATE $\bm{x}_{i+1} = \hat{\bm{x}}$\;
	        		\ELSE
	        			\STATE $\bm{x}_{i+1} =\bm{x}_{i}$\;
	        		\ENDIF
        		\ENDFOR
        		\STATE $\bm{\theta}_l^{(k+1)} \leftarrow \bm{x}_N$\;
        	\ENDFOR
        \STATE $k \leftarrow k+1$\;
	\ENDWHILE
	\STATE $m \leftarrow k-1$\;
	\STATE Compute model evidence estimate, $S \leftarrow \prod_{l=0}^{m} S_l$\;
	\STATE \bf for $l \leftarrow 1$ to $n$ do $\bm{\theta}_l = \bm{\theta}_l^{(m)}$
\end{algorithmic}
\end{algorithm}

For our implementation of TMCMC and the proposed GTMCMC (described next), the coefficient of variation\footnote{taken to be the sample standard deviation normalized by the sample mean} of the weights at any stage is a measure of the discrepancy between the intermediate PDFs associated with successive tempering parameter values, $\beta_k$ and $\beta_{k+1}$. Therefore, in order to control the rate of change in the intermediate PDFs, $\beta_{k+1}$ is chosen so that the resulting coefficient of variation of the weights is equal to some pre-specified value, $CoV$. This is essentially a root-finding problem and one could rely on the bisection method to determine $\beta_{k+1}$. As for the MCMC proposal covariance scaling parameter, a value of $\gamma^2=0.04$ is suggested in Ref.~\cite{ching2007transitional}. However, this is likely to be sub-optimal depending on the problem involved, particularly when parameters are highly correlated. In this work, we use an adaptive formulation for $\gamma^2$, first proposed in Ref.~\cite{catanach2017computational}, based on a feedback control strategy in order to produce a target MCMC acceptance rate of $0.234$. Though only one MCMC step, $N=1$, is taken per sample for all the results presented in Sections~\ref{sec:NumTests} and \ref{sec:LWD}, the algorithm allows for an arbitrary length for the chains in each stage in order to achieve convergence. TMCMC allows for a straightforward parallelization over the number of samples $n$ during each stage. During each stage, each sample will be advanced a number of MCMC iterations $N\geq 1$, and the sampling progresses in $m$ stages, resulting in a total of $n\times N\times m$ model evaluations. The $n$ model evaluations, with $n$ typically $10^3$ to $10^4$, can be performed in parallel at each stage and MCMC iteration, while the $N\times m$ model evaluations within chain and across stages are sequential in nature, with $N\approx$ 1 to 10 and $m\approx$ 10 to 100.

\section{Generalization of Transitional Markov chain Monte Carlo}
\label{sec:GTMCMC}

TMCMC is a sampling algorithm that relies on the concept of ``transitioning" an ensemble of realizations from an initial state, i.e. distributed according to a prior PDF in a Bayesian setting, to a final state, i.e. a posterior PDF. From an uncertainty point of view, the transition PDFs (from prior to posterior) are often associated with monotonically decreasing uncertainty. Therefore, TMCMC can be classified as an annealing algorithm \cite{neal1993probabilistic,fishman2013monte} that starts with an initial ``hot" state (with greater uncertainty) and ends at a final ``cold" solution (lower uncertainty). The samples are transitioned from one state to the next using sampling importance resampling (SIR) \cite{gelman2013bayesian}, whereby the resampling is achieved via multiple MCMC chains (one per unique sample) that aim to infuse diversity in the samples (ensuring that the ensemble is distributed according to the intermediate PDF associated with that particular stage).

Herein, we will present an extension, or generalization, of TMCMC that allows the algorithm to start with an arbitrary initial state, i.e. an importance PDF, $q(\bm{\theta})$, instead of the traditional prior PDF, $p(\bm{\theta})$. In some sense, the proposed technique is a tempering algorithm~\cite{marinari1992simulated}, rather than a stricter classification as an annealing algorithm, in that it generally allows both cooling and heating of the solution depending on the choice of $q(\bm{\theta})$. In this generalized framework, the modeler has the option of starting with an importance PDF exhibiting lower associated uncertainty in contrast to the posterior and let the algorithm transition from this ``colder" state to the target posterior. This might be helpful in practice when faced with implicit priors, for example, in which starting with a ``warmer" state might results with an initial ensemble with lots of samples in the infeasible region of parameter space (with zero associated prior probability). Furthermore, in the situation that the proposal PDF is chosen to be the prior one, the algorithm reverts back to TMCMC as will be described next.

Starting with Bayes' rule, we leverage the concept of importance sampling to introduce an importance PDF $q \left( \bm{\theta} \right)$ that we can easily sample from:
 \begin{align}
     p \left( \bm{\theta} \vert {\bm D} \right) & \propto q \left( \bm{\theta} \right) \frac{p \left(\bm{\theta} \right) p \left( {\bm D}  \vert \bm{\theta} \right)}{q \left(\bm{\theta} \right)} 
 \end{align}
This importance PDF will be used to initialize the proposed generalized TMCMC (GTMCMC) algorithm as follows. Similar to the approach first proposed by Neal \cite{neal2001annealed} and subsequently adopted for TMCMC \cite{beck2002bayesian,ching2007transitional}, we will sample from a series of `transitional’ or intermediate PDFs that are controlled by a tempering (or annealing) parameter, $\beta_m$, as in
 \begin{equation}
 	\begin{aligned}
 		p_j(\bm{\theta}) & \propto \left( \frac{p \left(\bm{\theta} \right) p \left( {\bm D}  \vert \bm{\theta} \right)}{q \left(\bm{\theta} \right)}  \right)^{\beta_j}q(\bm{\theta})\\
 		& = \left( p \left(\bm{\theta} \right) p \left( {\bm D}  \vert \bm{\theta} \right)  \right)^{\beta_j}q(\bm{\theta})^{1-\beta_j} \ ,
 	\end{aligned}
 	\label{eq:pjpdf}
\end{equation}
with $j = 0,\hdots,m$ and $0=\beta_0<\beta_1<\hdots<\beta_m=1$. As $\beta$ increases, the intermediate distribution $p_j(\bm{\theta})$ monotonically gets closer to the posterior in the KL-Divergence, see \ref{sec:info_mono}. As with traditional TMCMC, $\beta_j$ may be adaptively tuned using the coefficient of variation of the sample weights. The proof in \ref{sec:cov_mono} shows that the $CoV$ of the sample population is monotonically increasing so $\beta_j$ may be found efficiently using a variety of optimization methods.

The novelty of the proposed strategy lies in utilizing an importance distribution in the context of solving a Bayesian inverse problem, resulting in a new sequential monte Carlo sampler. Of most significant consequence is the potential to dramatically reduce the number of TMCMC stages, $m$, if an importance distribution, $q \left( \bm{\theta} \right)$, resembling the posterior PDF, $p \left( \bm{\theta} \vert {\bm D} \right)$, is used. In practice, the importance distribution is one that is easy to generate samples from, such as a multi-variate Gaussian. Although importance sampling can achieve remarkable results in drastically reducing the number of TMCMC stages, this can only be achieved when that distribution is chosen judiciously for each target posterior PDF. This will be demonstrated for an application of relevance to the oil and gas industry, whereby the proposed methodology significantly reduces the number of stages $m$ required in solving sequences of related Bayesian inference problems. More generally, the freedom to chose the initial density alleviates the limitation that TMCMC encounters when dealing with improper or complex prior PDFs that are difficult to sample from (the original TMCMC algorithm starts with initial samples that are obtained from Monte Carlo simulation of the pre-specified prior distribution). Note that if the importance density $q \left( \bm{\theta} \right)$ is chosen as the prior PDF $p \left( \bm{\theta} \right)$, then one reverts back to the standard TMCMC algorithm with the same intermediate PDFs. Therefore, this approach can be considered as a generalization of the original TMCMC sampling technique. The algorithm is summarized below:

\begin{algorithm}
	\footnotesize
	\caption{Generalized Transitional Markov chain Monte Carlo} 
	\label{gtmcmc}
	\begin{algorithmic}
	\REQUIRE Prior PDF, $p(\bm{\theta})$; Likelihood function, $p(\bm{D} \vert \bm{\theta})$; Importance PDF, $q(\bm{\theta})$; Number of samples, $n$; Target coefficient of variation, $COV$; Prescribed scaling factor for MCMC proposal covariance matrix, $\gamma$; MCMC chain length, $N$
	\ENSURE Samples from posterior PDF, $\{\bm{\theta}_1, \hdots, \bm{\theta}_n \}$ \& Model evidence estimate, $S$ 
	\STATE $\beta_0 \leftarrow 0$\; $k \leftarrow 0$\;
	\STATE Generate samples $\{\bm{\theta}_1^{(0)}, \hdots, \bm{\theta}_n^{(0)} \}$ from $q(\bm{\theta})$\;
	\WHILE{$\beta_k < 1$}
		\STATE Compute likelihood of ensemble of realizations $p(\bm{D} \vert \bm{\theta}_l^{(k)})$, $l = 1, \hdots, n$\;
		\STATE Compute the  weight of each sample $w(\bm{\theta}_l^{(k)})=\left( p \left(\bm{\theta}_l^{(k)} \right) p \left( {\bm D}  \vert \bm{\theta}_l^{(k)} \right)  \right)^{\beta_{k+1} - \beta_{k}}q(\bm{\theta}_l^{(k)})^{\beta_{k} - \beta_{k+1}}  $, with $\beta_{k+1}$ chosen such that coefficient of variation of those weights is equal to the target $COV$\;
		\STATE Compute contribution to model evidence, $S_k = \frac{1}{n}\sum_{l=1}^{n}w(\bm{\theta}_l^{(k)})$\;
		\STATE Compute the ensemble covariance matrix to be used in MCMC simulations:
		\begin{equation*}
			\tilde{w}(\bm{\theta}_l^{(k)}) = \frac{w(\bm{\theta}_l^{(k)})}{\sum_{l=1}^{n}w(\bm{\theta}_l^{(k)})} \ \ \ \ \ \ \ \overline{\bm{\theta}}^{(k)} = \sum_{k=1}^{n}\bm{\theta}_l^{(k)} \tilde{w}(\bm{\theta}_l^{(k)})
		\end{equation*}
		\begin{equation*}
			\bm{\Sigma}^{(k)} = \gamma^2\sum_{k=1}^{n}\tilde{w}(\bm{\theta}_l^{(k)})(\bm{\theta}_l^{(k)} - \overline{\bm{\theta}}^{(k)})(\bm{\theta}_l^{(k)} - \overline{\bm{\theta}}^{(k)})^T
		\end{equation*}\\
		\STATE Resampling: Generate $n$ random samples, $\{\tilde{\bm{\theta}}_1^{(k)}, \hdots, \tilde{\bm{\theta}}_n^{(k)} \}$, from the current ensemble $\{\bm{\theta}_1^{(k)}, \hdots, \bm{\theta}_n^{(k)} \}$ with corresponding probabilities $\tilde{w}(\bm{\theta}_l^{(k)})$\;
		\FOR{$l\leftarrow 1$ \textbf{to} $n$}
			
			\STATE $\bm{x}_0 \leftarrow \tilde{\bm{\theta}}_l^{(k)}$\;
			\FOR{$i\leftarrow 0$ \textbf{to} $N-1$}
				\STATE Generate $\hat{\bm{x}}$ from Gaussian proposal $\mathcal{N}(\bm{i}_k, \bm{\Sigma}^{(k)})$\;
				\STATE Compute acceptance probability, $\alpha = {\rm min} \left\{1, \frac{\left( p \left(\hat{\bm{x}} \right) p \left( {\bm D}  \vert \hat{\bm{x}} \right)  \right)^{\beta_{k+1}}q(\hat{\bm{x}})^{1-\beta_{k+1}}}{\left(p \left(\bm{x}_{i} \right) p \left( {\bm D}  \vert \bm{x}_{i} \right)  \right)^{\beta_{k+1}}q(\bm{x}_{i})^{1-\beta_{k+1}}}\right\}$\;
				\STATE Generate $u$ from from the continuous uniform distribution on the interval  $\left[0,1\right]$\;
				\IF{$u < \alpha$}
					\STATE $\bm{x}_{i+1} = \hat{\bm{x}}$\;
				\ELSE
					\STATE $\bm{x}_{i+1} =\bm{x}_{i}$\;
				\ENDIF
			\ENDFOR
			\STATE $\bm{\theta}_l^{(k+1)} \leftarrow \bm{x}_N$\;
		\ENDFOR
		\STATE $k \leftarrow k+1$\;
	\ENDWHILE
	\STATE $m \leftarrow k-1$\;
	\STATE Compute model evidence estimate, $S \leftarrow \prod_{l=0}^{m} S_l$\;
	\STATE \bf for $l \leftarrow 1$ to $n$ do $\bm{\theta}_l = \bm{\theta}_l^{(m)}$
\end{algorithmic}
\end{algorithm}

As was mentioned previously for TMCMC, at any stage $\beta_{k+1}$ is chosen so that the resulting coefficient of variation of the weights is equal to some pre-specified value, $\mathrm{CoV}$, which again could be obtained through the bisector method or any other appropriate root-finding method. Similar to our implementation of standard TMCMC, we rely on an adaptive formulation for determining $\gamma^2$ in order to produce an optimal MCMC acceptance (see \cite{catanach2017computational} for more details). It is important to note that the GTMCMC algorithm as presented relies on the Metropolis Hastings MCMC at every stage, but that can be replaced by a more efficient MCMC sampler for a given inverse problem.

\section{Numerical Tests}
\label{sec:NumTests}
In this section, we will demonstrate the advantages of the GTMCMC algorithm over the TMCMC algorithm for a series of canonical test problems. In section \ref{sec:4DGaussian}, we generate samples from a 4-dimensional PDF using the TMCMC and the GTMCMC algorithm with different importance distributions and compare their errors in the mean prediction and the model evidence. In section \ref{sec:scalability}, we perform scalability studies on a series of inverse problems of dimensionalities ranging from 1 through 8. In section \ref{bimodalGaussian}, we evaluate the performance of the GTMCMC algorithm in estimating the model evidence with associated bi-modal posterior distributions. In section \ref{sec:rosenbrock}, we examine the performance of the proposed algorithm in sampling a strongly non-Gaussian posterior density with an associated negative log-likelihood Rosenbrock function.

\subsection{Test 1: 4D Gaussian Prior and Likelihood}
\label{sec:4DGaussian}
We first show the impact of using different importance distributions for the GTMCMC algorithm by solving a Bayesian inference problem with 4-dimensional (4D) standard Normal likelihood function. The prior distribution is a 4D tensor product of 1D Gaussian distributions all with mean and standard deviation of $(\mu,\sigma)=(1,5)$, resulting in a Gaussian posterior distribution with associated mean of $1/26$ and variance of $25/26$ for each parameter. The mean of the importance PDF for the GTMCMC algorithm is fixed to $1$ in all dimensions while the variance takes the values $0.6^2$, $1^2$, $3^2$, and $5^2$, independently for each dimension. Note that the last value for the variance coincides with that of the prior PDF, allowing us to also investigate the performance of the TMCMC algorithm. We also setup a test case with an importance PDF equal to the posterior distribution. We run the algorithm for each importance distribution with the hyper-parameter CoV varying from 0.1 to 1.

\begin{figure}[!ht]
    \centering
    \includegraphics[width=0.5\columnwidth]{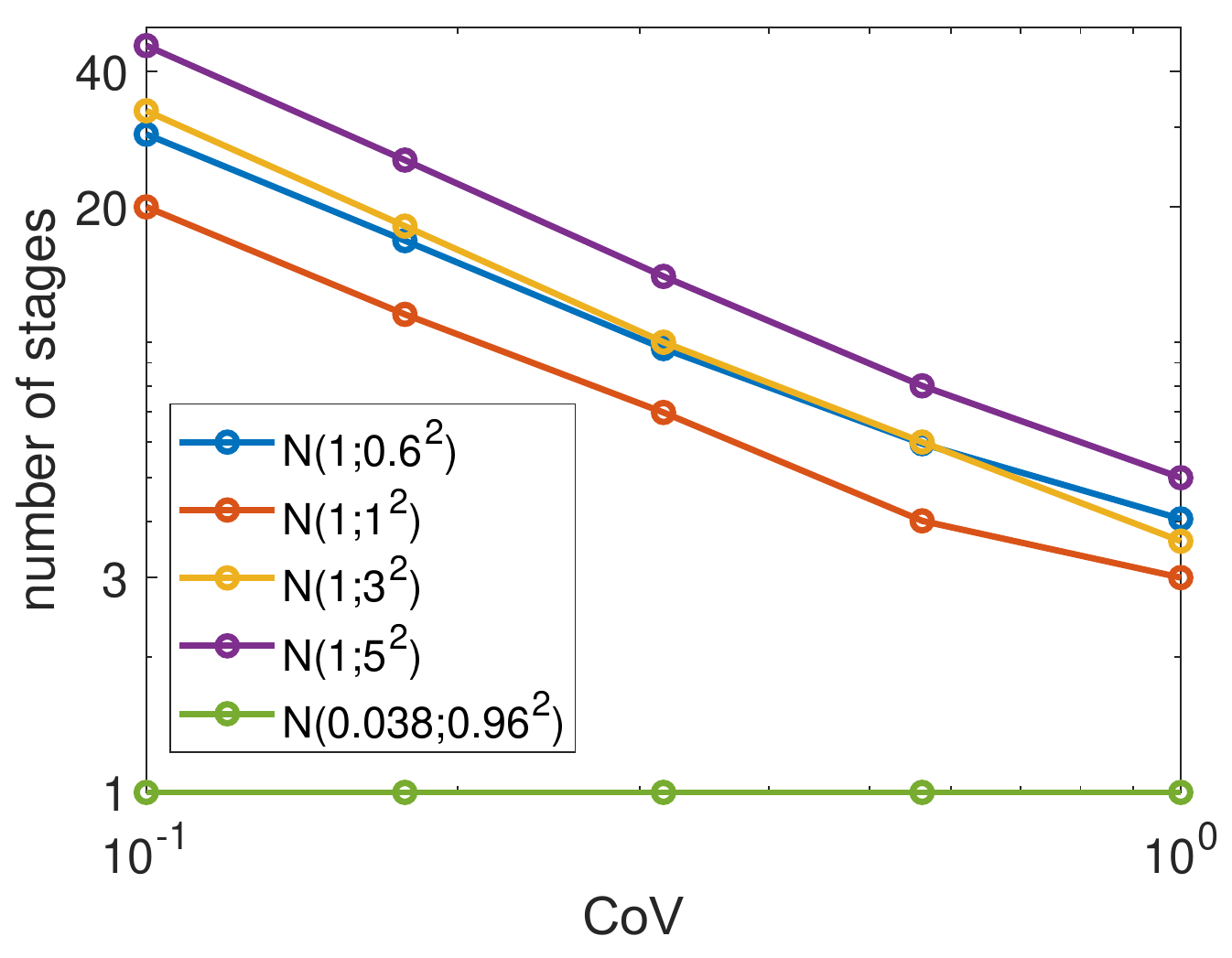}
    \caption{The average number of GTMCMC stages vs. CoV for different importance distributions 
    }
    \label{fig:impact_of_proposal}
\end{figure}

For the following set of results, GTMCMC is executed 1000 times to extract the average performance of this stochastic sampling algorithm. The number of samples per stage is fixed to $5\times10^3$ for all tests. Figure~\ref{fig:impact_of_proposal} shows the average number of stages used by the algorithm for each configuration across the 1000 repeated runs. The coefficient of variation, CoV, controls the evolution of tempering parameter $\beta$. A smaller CoV induces more transition stages to reach $\beta=1$ and yields a more accurate result. We observe a near linear decrease in the number of TMCMC stages as we increase the allowable coefficient of variation. In the case that the importance distribution is chosen to be the posterior distribution (available analytically for this problem), the GTMCMC algorithm converges in one step independent of the CoV value as expected since the initial set of samples are already drawn according to the target posterior. For the case in which the importance distribution is chosen to be the prior distribution, the GTMCMC algorithm reverts back to TMCMC, requiring a larger number of intermediate PDFs (and stages) to reach convergence in comparison to more compact importance distributions.

\begin{figure}[!ht]
	\centering
	\subfigure[]{
		\begin{minipage}[t]{0.45\linewidth}
			\centering
			\includegraphics[width=\columnwidth]{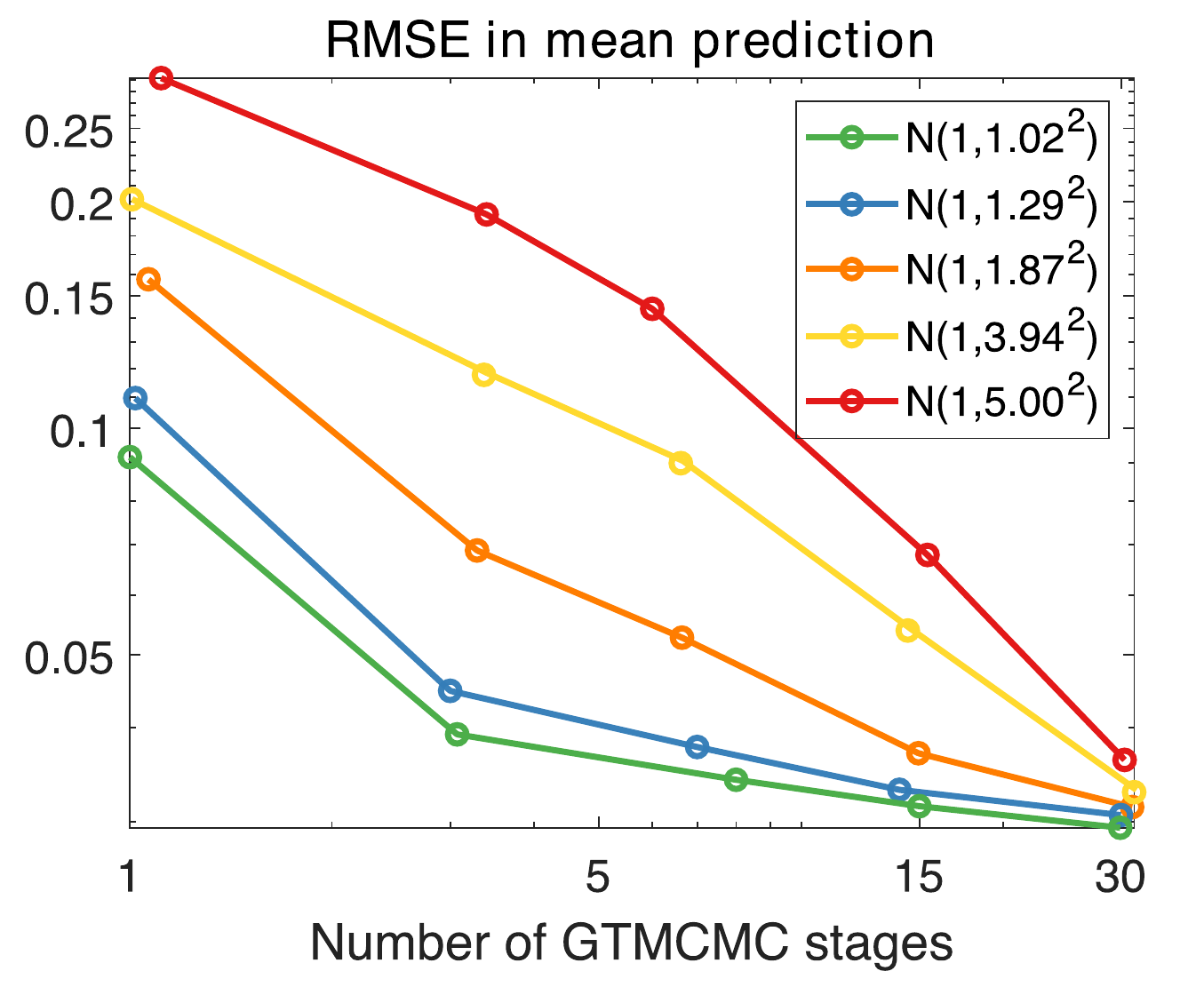}
		\end{minipage}%
	}%
	\subfigure[]{
		\begin{minipage}[t]{0.45\linewidth}
			\centering
			\includegraphics[width=\columnwidth]{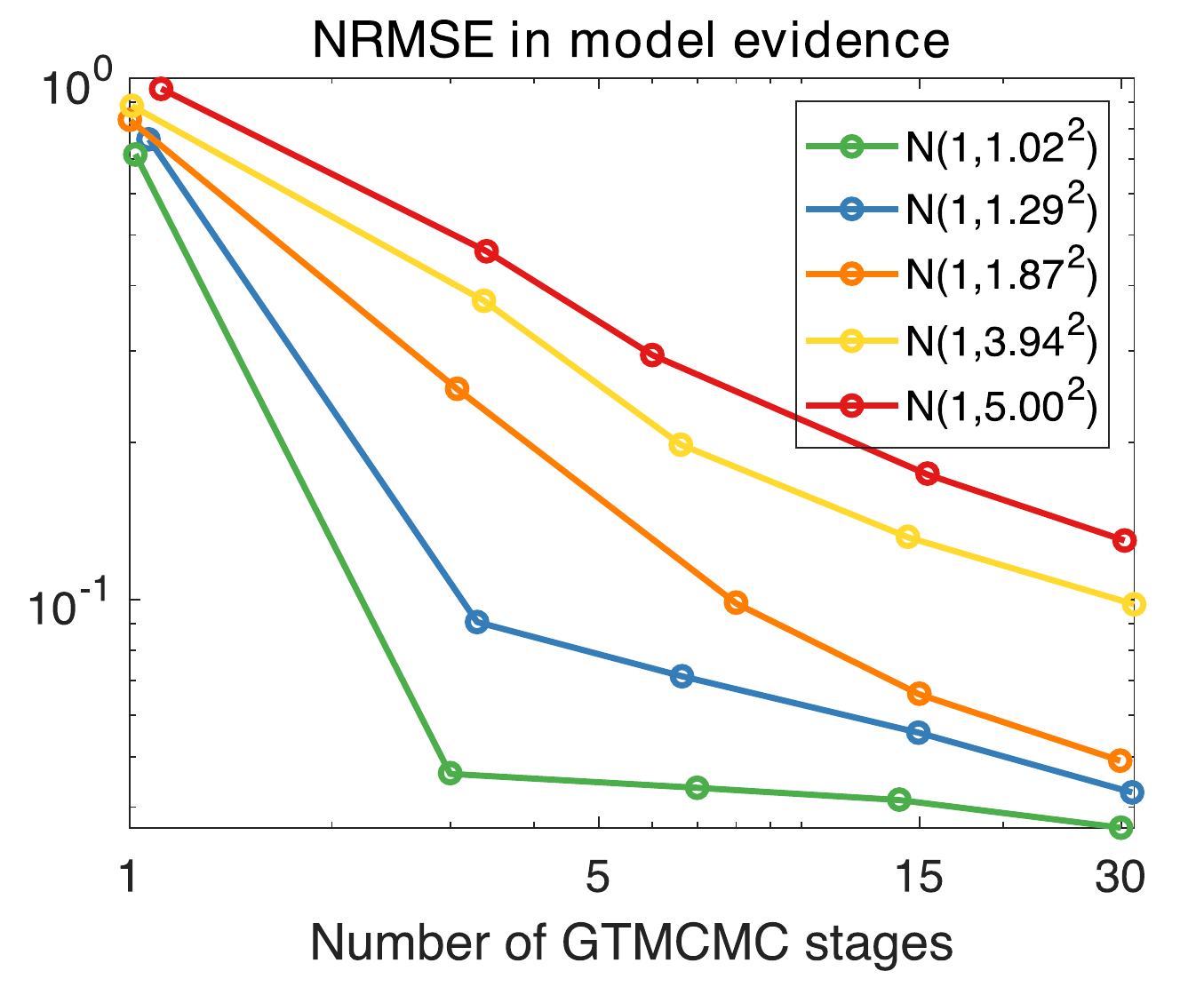}
		\end{minipage}%
	}
	\caption{The error in estimating the posterior parameter mean and model evidence using GTMCMC versus the number of stages allowed: (a) RMSE in the posterior mean parameter estimates and (b) NRMSE in the evidence estimates. The curve labels refer the importance marginal distribution for each parameter.
	}
	\label{fig:compare_proposals}
\end{figure}

Next, the effect of importance distribution variance on the performance of GTMCMC is examined. For a fair comparison, the CoV is adjusted to ensure that the algorithm takes a target number of stages. The number of samples in each stage is $5\times 10^3$ and again the algorithm is run 1000 times for each test. Figure~\ref{fig:compare_proposals} compares the root-mean-square-error (RMSE) in the posterior mean estimate and the normalized\footnote{Normalization is achieved with respect to the corresponding analytical values of the model evidence.} RMSE (NRMSE) in the model evidence estimate obtained using the various importance distributions.

For all experiments, both the posterior mean estimates and the evidence estimates increase in accuracy as the number of stages increases (controlled via smaller CoV values). The errors associated with the standard TMCMC algorithm (red curves) indicate a diminished performance in comparison to GTMCMC relying on importance PDFs with lower associated variance than the prior PDF. 

These results suggest that given a computational budget based on the number of stages, the GTMCMC algorithm provides estimates of greater accuracy with a choice for importance distribution that is "closer" to the target posterior. Furthermore, the rate of decrease in the errors is greater for GTMCMC with certain importance distributions. Therefore, a judiciously chosen importance distribution can potentially result in computational savings for a given error budget associated with a threshold error. It is important to note that an importance distribution with greater uncertainty than that associated with the prior PDF may diminish the performance of GTMCMC in comparison to standard TMCMC .

The error in the model evidence estimates, shown in Figure \ref{fig:compare_proposals}(b), exhibits similar trends as those in the posterior mean estimates. For one allowable stage, all importance distributions result in similar NRMSE values. For greater number of stages examined, GTMCMC outperforms TMCMC in reducing the NRMSE in model evidence estimates for the importance distributions investigated

These observations suggest, as expected, that an importance PDF that is "closer" to the target posterior provides more precise estimates of the posterior mean and model evidence. They also suggest that in practice it might be deemed worthwhile to focus on obtaining an "optimal" choice for the importance PDF, one that minimizes the error given a computational budget (number of allowable stages). This aspect is beyond the scope of this investigation.

\subsection{Test 2: Performance with increased dimensionality}
\label{sec:scalability}

\begin{figure}[!ht]
    \centering
    \subfigure[]{
    \begin{minipage}[t]{0.45\linewidth}
    \centering
    \includegraphics[width=\columnwidth]{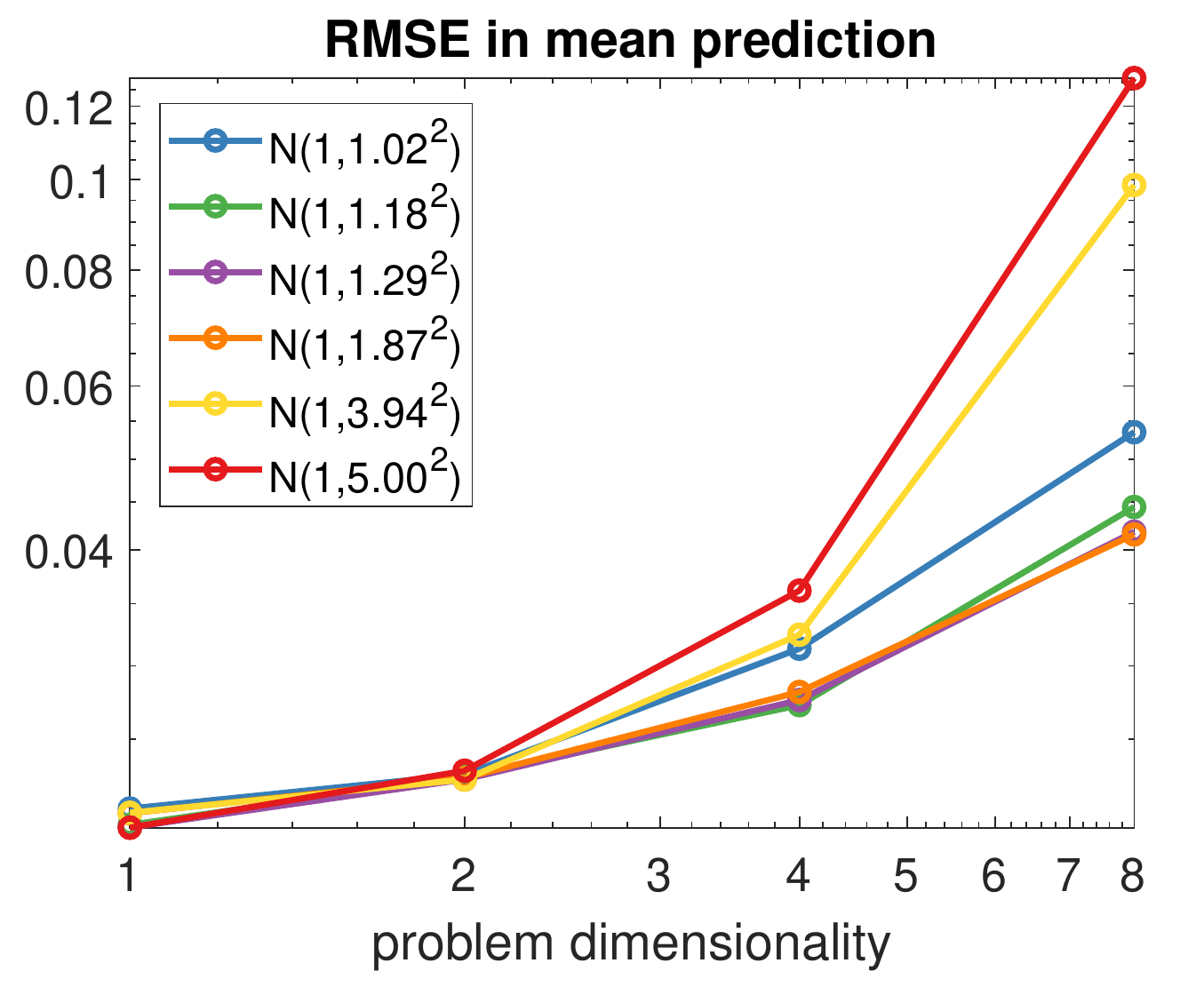}
    \end{minipage}%
    }%
    \subfigure[]{
    \begin{minipage}[t]{0.45\linewidth}
    \centering
    \includegraphics[width=\columnwidth]{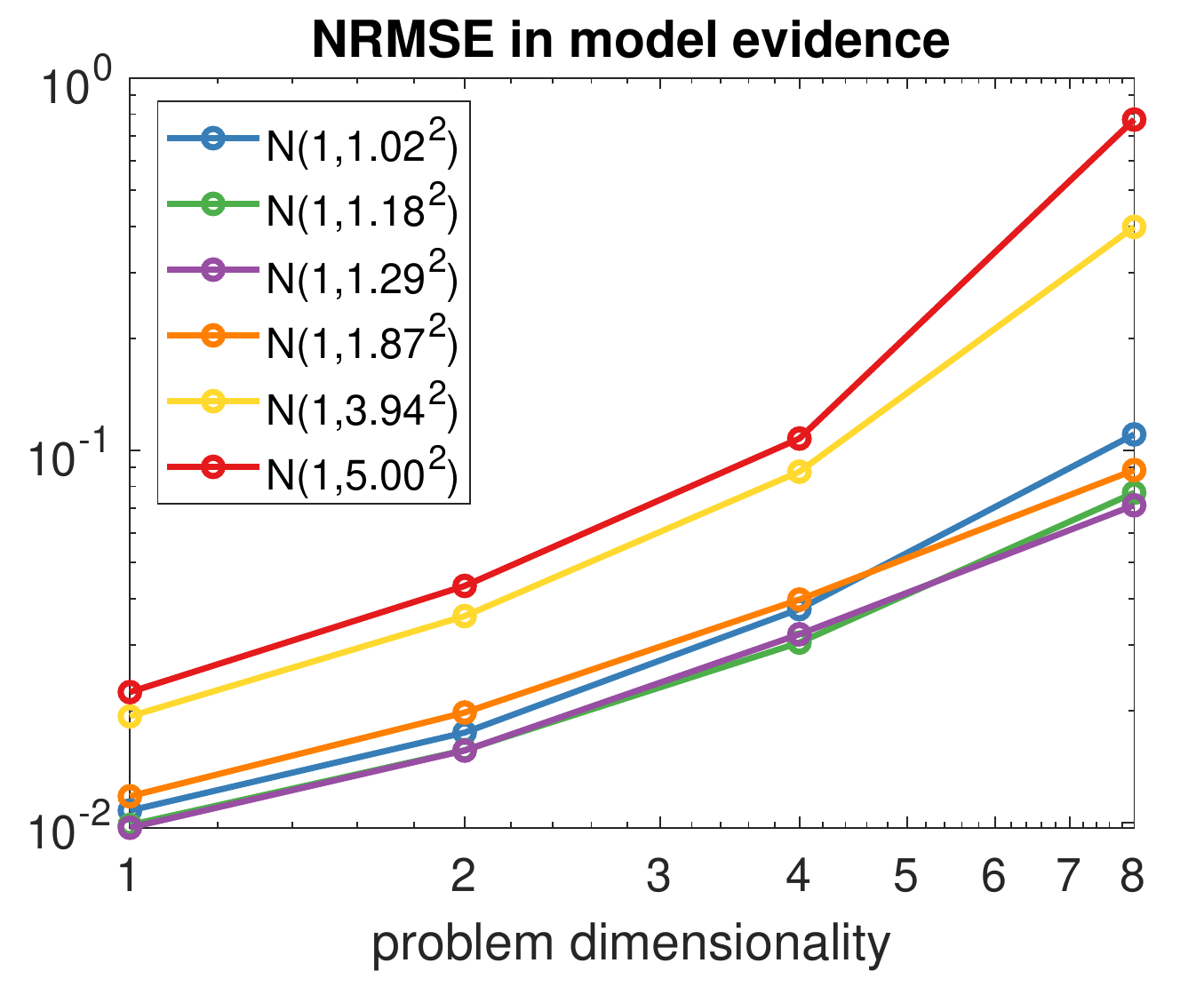}
    \end{minipage}%
    }
    \caption{(a). The RMSE in the mean prediction and (b) The NRMSE in the evidence estimation when increasing problem dimensionality.}
    \label{fig:dim_scalability}
\end{figure}

For the following investigation, the number of TMCMC stages is fixed to 20 in order to isolate the effect of the problem dimensionality (number of parameters to be inferred) in estimating the posterior parameter mean and model evidence. The same set of importance distributions are employed as in the previous subsection. The likelihood is a tensor product of one-dimensional standard Normal PDFs with problem dimensionality increasing from 1 to 8. The prior distribution for all experiments is a tensor product of one-dimensional Gaussians $\textbf{N}(1,5^2)$, resulting in a marginal posterior distribution of $\textbf{N}(0.0385,0.9625^2)$ for all parameters. 
The number of samples used for GTMCMC is fixed at $5\times10^3$. Error statistics are based on $2\times 10^3$ repeated runs. Figure~\ref{fig:dim_scalability} shows the error in the posterior mean and model evidence estimates for different problem dimensionalities. It is obvious that the error in the posterior mean and the evidence estimates increase with the problem dimensionality, as expected (with a fixed number of samples and CoV). For problem dimensionalities greater than one, GTMCMC outperforms TMCMC for all chosen importance PDFs with smaller associated estimation errors than those obtained with TMCMC (GTMCMC with an importance distribution equal to the prior).

\subsection{Test 3: Bi-modal Gaussian Posterior}
\label{bimodalGaussian}
\begin{figure}[!ht]
    \centering
    \includegraphics[width=0.5\columnwidth]{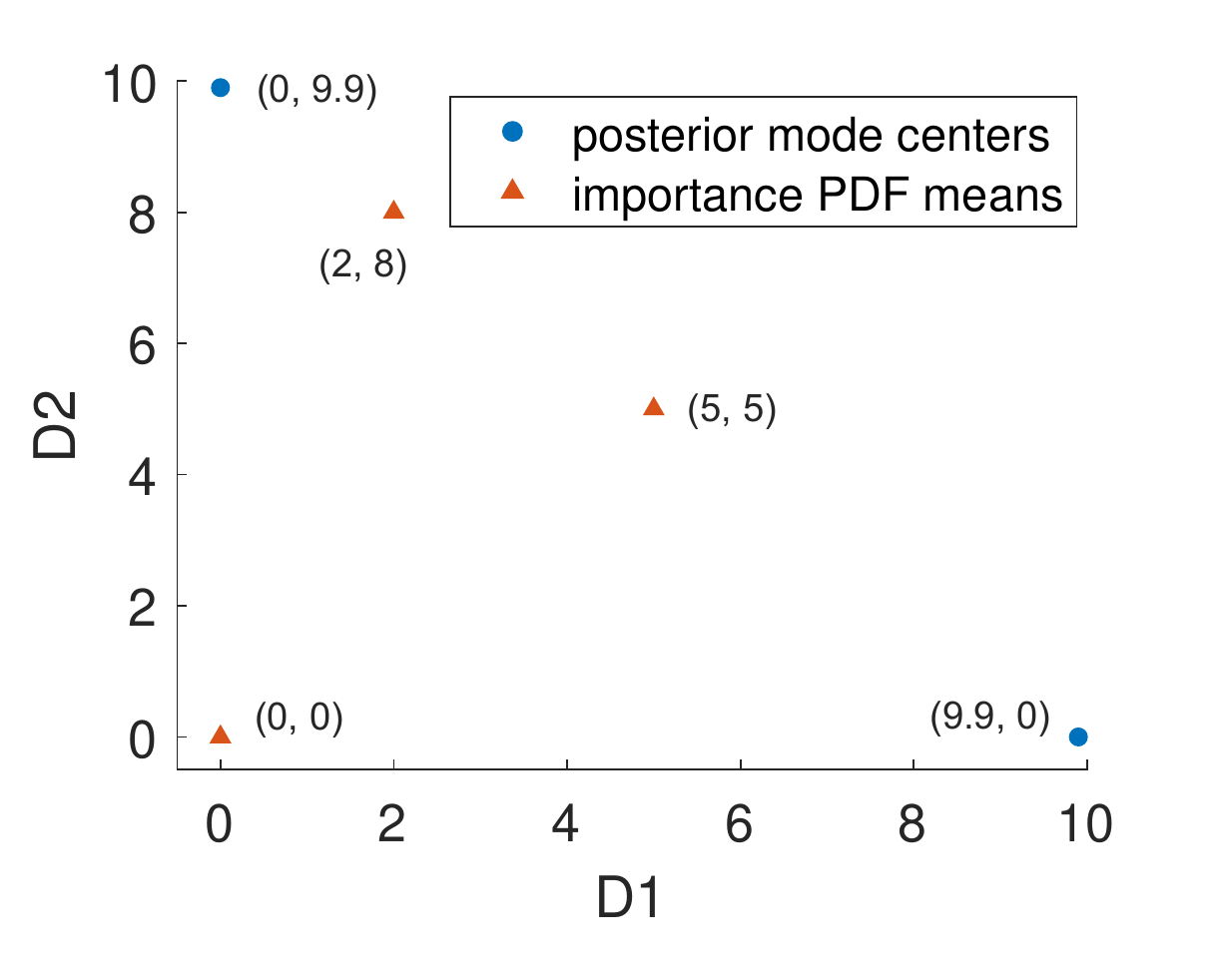}
    \caption{The centers of two posterior modes and selected importance distribution means. }
    \label{fig:bimodal_centers}
\end{figure}

In this subsection, we evaluate the performance of the GTMCMC algorithm for sampling a canonical 2D bi-modal posterior distribution. The likelihood is constructed as a weighted sum of two Gaussian distributions with parameters shown in Table~\ref{table:bimodalParam}, where $\textbf{I}_n$ represents the $n$-D identity matrix. With a multi-variate normal prior distribution centered at $((0,0)$ and diagonal covariance matrix $10^2\times \textbf{I}_2$, the posterior distribution is computed analytically with parameters given in Table~\ref{table:bimodalParam}. We run the GTMCMC algorithm with Gaussian importance distributions with several sets of parameter values. The posterior modes and selected importance means are shown in Figure~\ref{fig:bimodal_centers}. We set the covariance matrix to $10^2\times \textbf{I}_2$ for all mean values, and the test with the mean value of $(0, 0)$ represents the standard TMCMC algorithm. We adjust the CoV hyper-parameter to make the algorithm converge in 5, 10, 20, and 30 stages, respectively. Each configuration is simulated $10^4$ times, and each run uses $10^4$ samples. 
\begin{table}[!ht]
\centering
\captionsetup{justification=centering}
\caption{The likelihood and posterior parameters.}
\begin{tabular}{ccccc}
\toprule
 & & center & cov. matrix & weight\\
\midrule
\multirow{2}*{Likelihood}&Peak 1& (10, 0)&$\textbf{I}_2$&25\% \\
&Peak 2& (0, 10)&$\textbf{I}_2$&75\% \\
\multirow{2}*{Posterior}&Peak 1& (9.9, 0)&$0.99\times \textbf{I}_2$&25\% \\
&Peak 2& (0, 9.9)&$0.99\times \textbf{I}_2$&75\% \\
\bottomrule
\end{tabular}
\label{table:bimodalParam}
\end{table}

\begin{figure}[!ht]
    \centering
    \includegraphics[width=0.5\columnwidth]{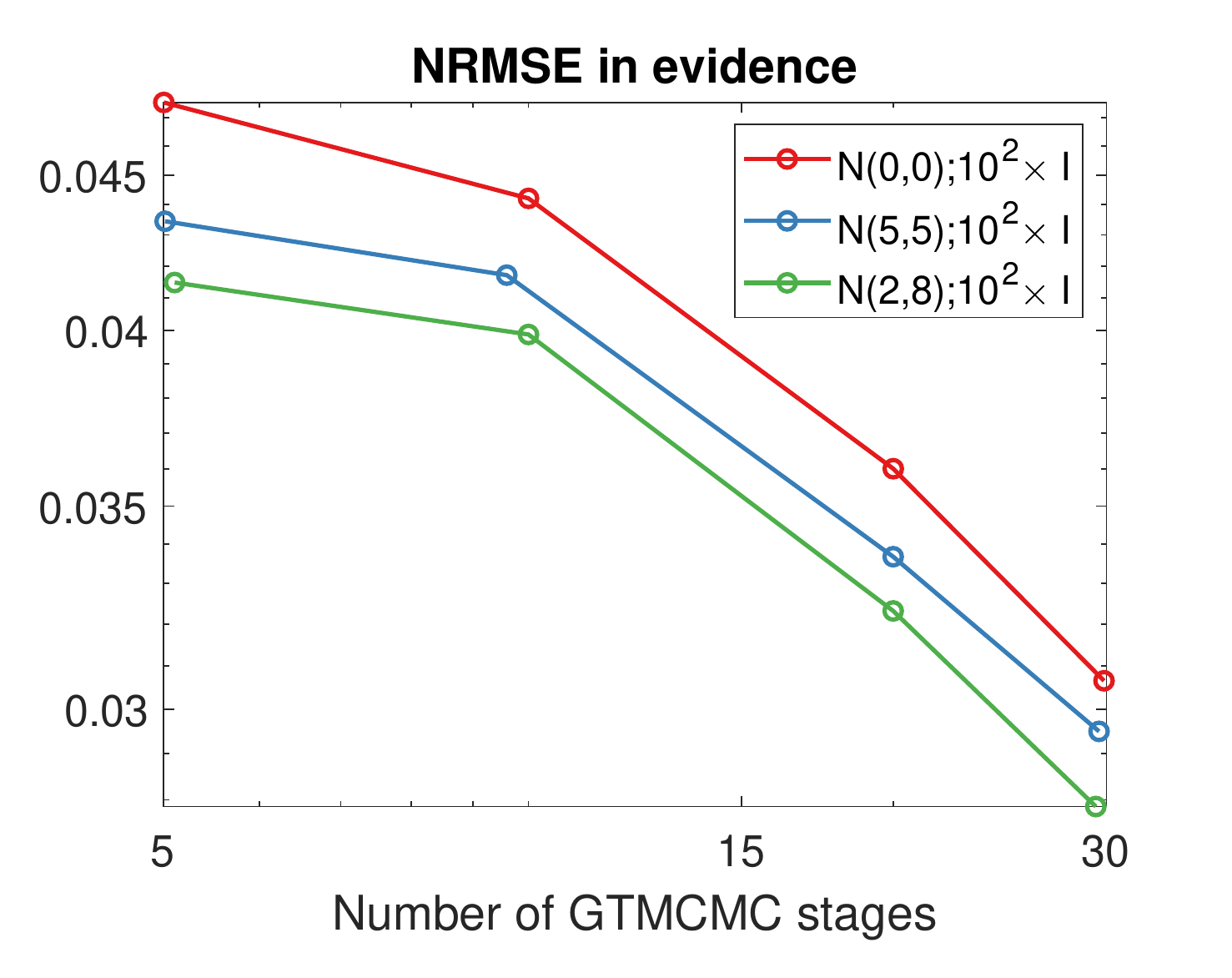}
    \caption{The NRMSE in the model evidence estimation.}
    \label{fig:bimodal_result}
\end{figure}

%

Figure~\ref{fig:bimodal_result} shows the error in evidence estimation.
Similar to results shown in Figure \ref{fig:compare_proposals}(b), as more GTMCMC stages are included, the error in the model evidence decreases. Meanwhile, GTMCMC with importance PDFs centered at $(5, 5)$ and $(2, 8)$ always produce smaller error than the standard TMCMC algorithm (the red curve). 

\subsection{Test4: 3D Rosenbrock Posterior}
\label{sec:rosenbrock}
\begin{figure}[!ht]
    \centering
    \subfigure[TMCMC]{
        \begin{minipage}[t]{1\linewidth}
        \centering
        \includegraphics[width=\columnwidth]{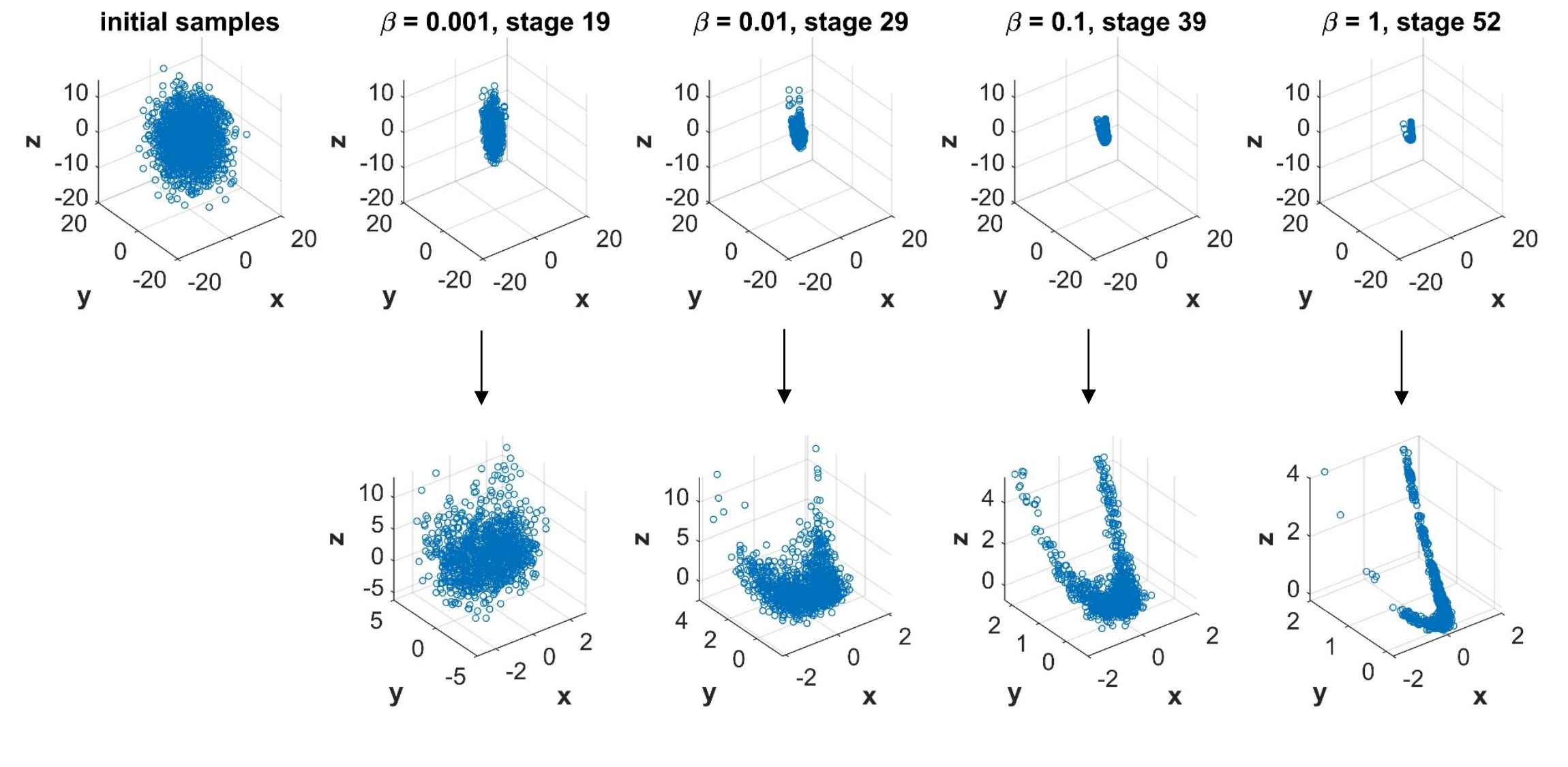}
        \end{minipage}
    }
    
    \subfigure[GTMCMC]{
        \begin{minipage}[t]{1\linewidth}
        \centering
        \includegraphics[width=\columnwidth]{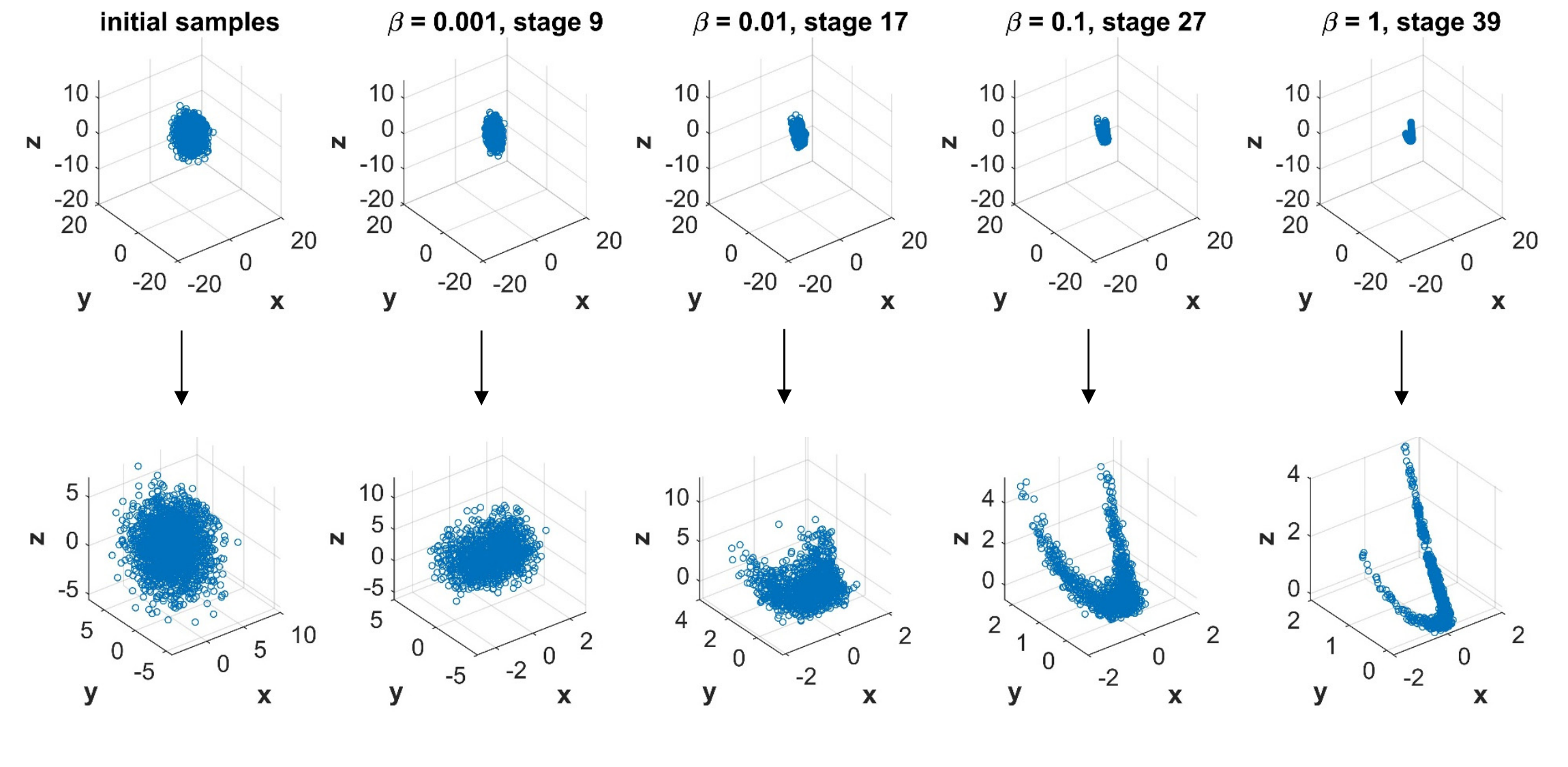}
        \end{minipage}
    }

    \caption{3D scatter plots of samples from different stages of the (a) TMCMC and (b) GTMCMC algorithms. The zoomed-in figures show the details of the sample distribution in different stages.}
    \label{fig:rosenbrockSamples}
\end{figure}

In this subsection, we monitor the evolution of samples from the GTMCMC algorithm. The log-likelihood of the tested problem is defined is as follows
\begin{equation}
    \centering
    \text{log}(p(d\vert x, y, z))\propto -(100\times(y-x^2)^2 + (1-x)^2 + 100\times(z-y^2)^2 + (1-y)^2)
\end{equation}
which is a 3D extension of the Rosenbrock function with variables $x, y,$ and $z$. The maximum likelihood is inside a long, narrow, parabolic shaped valley in the 3D space. We set the prior distribution to be N$((0,0,0);5^2\times \textbf{I}_3)$. For the GTMCMC algorithm, we choose the importance distribution to be N$((1,1,1);2^2\times \textbf{I}_3)$ which is more informative than the prior distribution. We run the standard TMCMC and the GTMCMC algorithms for this problem with $2\times 10^3$ samples and $\mathrm{CoV}=0.2$ for both. It takes $52$ stages for TMCMC to transition from prior to posterior while for GTMCMC the number of stages is $39$. Figure~\ref{fig:rosenbrockSamples} shows 3D scatter plots of samples from different stages that correspond to select $\beta$ values. The bottom rows of Figures~\ref{fig:rosenbrockSamples}(a) and \ref{fig:rosenbrockSamples}(b), respectively, show zoomed-in regions where samples are concentrated. The initial samples in the GTMCMC algorithm are more concentrated since the importance distribution of the algorithm has a smaller variance than the prior distribution. As a result, the GTMCMC algorithm takes less stages to converge compared to the standard TMCMC algorithm.

\section{Inversion of Ultra-Deep Directional Resistivity Measurements}
\label{sec:LWD}

In this section, we investigate the performance of the GTMCMC algorithm and compare its results with those obtained from the standard TMCMC algorithm in the inversion of directional resistivity measurements. The inversion and interpretation of well-logging data play a crucial rule in real-time geosteering as it steers the well in a way that maximizes the contact with the pay zone~\cite{lesso1996principles}. Among several well-logging technologies, azimuthal resistivity logging-while-drilling (LWD) is widely used due to its relatively large depth of detection and azimuthal directional sensitivity~\cite{bittar2000electromagnetic}. The azimuthal resistivity well-logging data is challenging because the logging tool responses are usually nonlinear to formation topology. As the detection range of the logging tool grows, more parameters are needed to describe the underground formation of interest, which makes these inverse problems more challenging~\cite{key20091d,beer2010geosteering}. 

Deterministic optimization methods such as Newton-based methods, are widely used in the industry~\cite{habashy2004general,beer2010geosteering} due to their computational scalability. However, gradient-based approaches can only find local optima, with the result heavily depending  on the initial guess. Even though one can apply multi-start methods~\cite{marti2003multi} to increase the chances of finding the global optimal solution, deterministic optimization methods cannot provide associated uncertainty caused by noise in the data, which is integral to well-logging inverse problems. Therefore, stochastic methods based on Bayesian theory have drawn great attention in solving well-logging related problems in recent years. In Ref.~\cite{kullawan2014decision} a decision-analytic framework to support high-quality geosteering decisions is implemented based on Bayesian inference. In Refs.~\cite{shen2018solving} and \cite{lu2019parallel} the authors proposed to use MCMC methods to obtain formation parameters of 1D earth models from synthetic directional resistivity well-logging data. Instead of inferring formation parameters of fixed dimensionality, the authors in  \cite{minsley2011trans, sambridge2013transdimensional,shen2020parallel} used trans-dimensional MCMC methods to infer parameters of earth models that can have a varied number of layers. Miao {\em et al.}~\cite{miao2020nonlinear} and Veettil {\em et al.}~\cite{veettil2020bayesian} proposed to infer well-bore locations from gamma-ray (GR) logging data using Bayesian approaches.  In this work, we solve the well-logging inverse problems using resistivity EM data with the GTMCMC algorithm, and compare its results with those obtained from the standard TMCMC method.

\subsection{Ultra-Deep Directional Resistivity Measurements}
\label{sec:LWD_problem}
\begin{figure}[!ht]
    \centering
    \includegraphics[width=0.5\columnwidth]{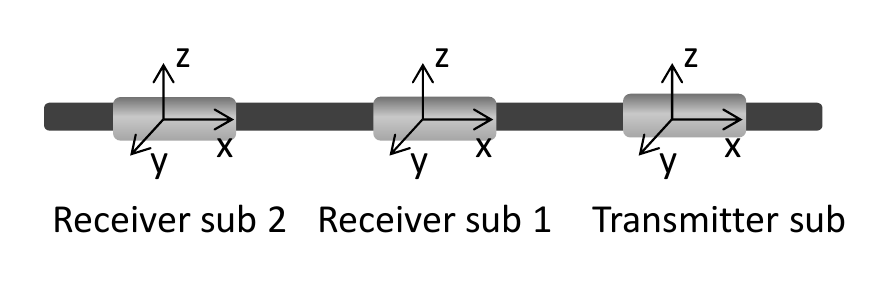}
    \caption{Schematic diagram of a directional EM LWD tool. The tool is configured with 1 transmitter and 2 receiver antennas.}
    \label{fig:loggingTool}
\end{figure}

The ultra-deep directional logging tool adopts multiple pairs of transmitter-receiver subs configured with different spacing and frequency. Figure \ref{fig:loggingTool} demonstrates a schematic diagram of a directional resistivity logging tool with 1 transmitter and 2 receivers. The tri-axial antennas are oriented at \textit{x}-, \textit{y}- and \textit{z}-directions respectively. The working frequency can vary from one kHz to several hundreds kHz, thus enabling a large detection depth and a wide range of resistivity. Instead of directly using the tensor voltage, attenuation and phase shift signals delivered by the logging tool are used for model inversion~\cite{seydoux2014full}. 

In this work, we design a synthetic ultra-deep directional logging tool configured with 1 transmitter sub and 3 receiver subs. Three working frequencies, 2 kHz, 6 kHz and 24 kHz are available to three transmitter-receiver pairs. For each combination of spacing and frequency in a single channel, four types of phase shift and attenuation measurements are constructed based on the formation coupling tensors~\cite{li2005new,dong2015application}. The four types of measurements provide sensitivity to different geological features including the formation boundary, relative dip angle, bulk resistivity, and anisotropy respectively. As a result, $72$ measurements are generated at each logging station, which corresponds to one unique set of model parameters. In total, there are 14400 measurements generated for 200 logging stations. We consider several noise levels for the phase shift (ps) and the attenuation signals (att). These are denoted as $\sigma_{ps}$ and $\sigma_{att}$ respectively.

\subsection{Inversion of Ultra-Deep Directional Resistivity Measurements}
\label{sec:LWD_inversion}

\begin{figure}[!ht]
    \centering
    \includegraphics[width=0.45\columnwidth]{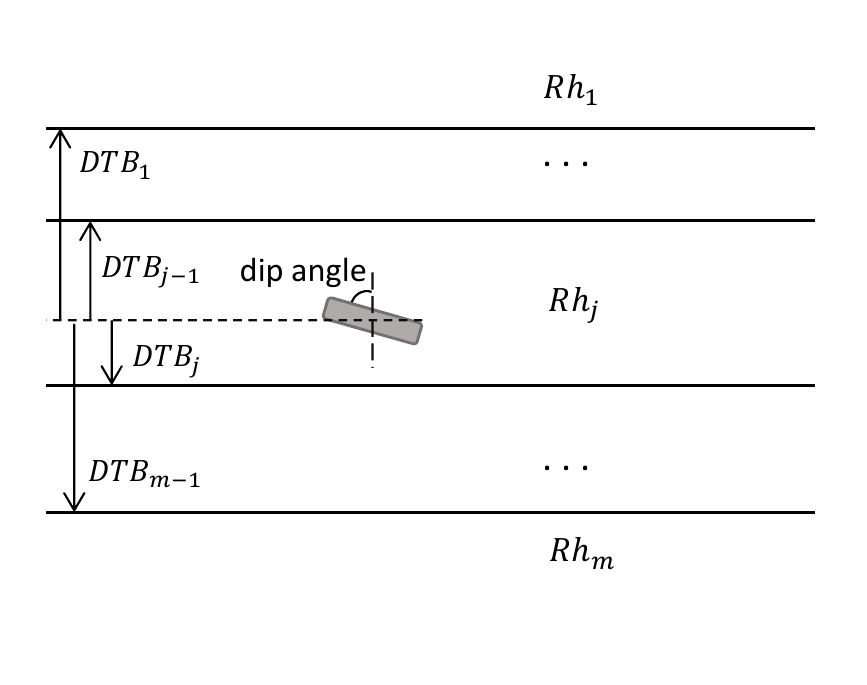}
    \caption{Parameters in an 1D earth model. The dip angle is considered to be known during the drilling process. }
    \label{fig:probDef}
\end{figure}

The parameters of an 1D earth model are shown in Figure \ref{fig:probDef}. Here, Rh$_j$ represents the horizontal resistivity of the $j$-th layer, and DTB$_j$ represents the distance from the LWD tool to the lower boundary of the $j$-th layer. A positive DTB value indicates that the boundary is below the logging tool and vice versa, and DTB$_i >$ DTB$_j$ if $i >j$. To make sure that the sampling algorithm always generates physically-meaningful DTB parameters, we transform DTB$_j$ to thickness parameters thk$_j > 0$ by 
\begin{equation}
\mathrm{thk}_j = \mathrm{DTB}_j - \mathrm{DTB}_{j-1}, \quad j=2,3,...,m-1
\end{equation}
As mentioned before, we also infer two noise standard deviations. Consequently, the inversion algorithm infers $(\mathrm{Rh}_1, \hdots,\mathrm{Rh}_m,\mathrm{DTB}_1,\mathrm{thk}_2,\hdots,\mathrm{thk}_{m-1},\sigma_{ps},\sigma_{att})$ for an $m$-layer 1D earth model. Considering the detection scope of the ultra-deep directional logging tool, the tested earth models have at most 7 layers in this investigation. 

The LWD inverse problem can also be solved through pixel-based inversion~\cite{lin2012schemes,thiel2017high} that discretizes the underground formation into many thin layers so that one needs to infer only Rh for each thin layer to reconstruct the earth model. Pixel-based inversion offers more flexibility since it does not predetermine the number of layers, however, it involves dozens to hundreds of unknown parameters, which is impractical to be solved by a stochastic inverse algorithm in real-time. Therefore, we will focus on solving model-based LWD inverse problems in this paper.    

The inversion for the ultra-deep directional resistivity measurements requires numerous forward simulations, which is the most computationally intensive part in the inverse problem. TMCMC computes forward simulations for a set of samples independently in each stage, and it is therefore naturally parallelizable as opposed to most of the other MCMC methods. The model evaluations are performed in parallel via MPI~\cite{gropp1999using} for efficient computations on a cluster with distributed computer cores~\cite{safta2020transitional}. 

\subsection{Experimental Results}
\label{sec:LWD_results}

Given the subsurface structure is more or less similar at different logging points, we are able to incorporate information from the previous logging point into the current logging point with the GTMCMC algorithm. In this subsection, we show that the GTMCMC algorithm can improve the result of LWD inverse problems in a 2D earth model with much less computational cost, comparing to the standard TMCMC algorithm. 

\begin{figure}[!ht]
\centering
\includegraphics[width=0.8\columnwidth]{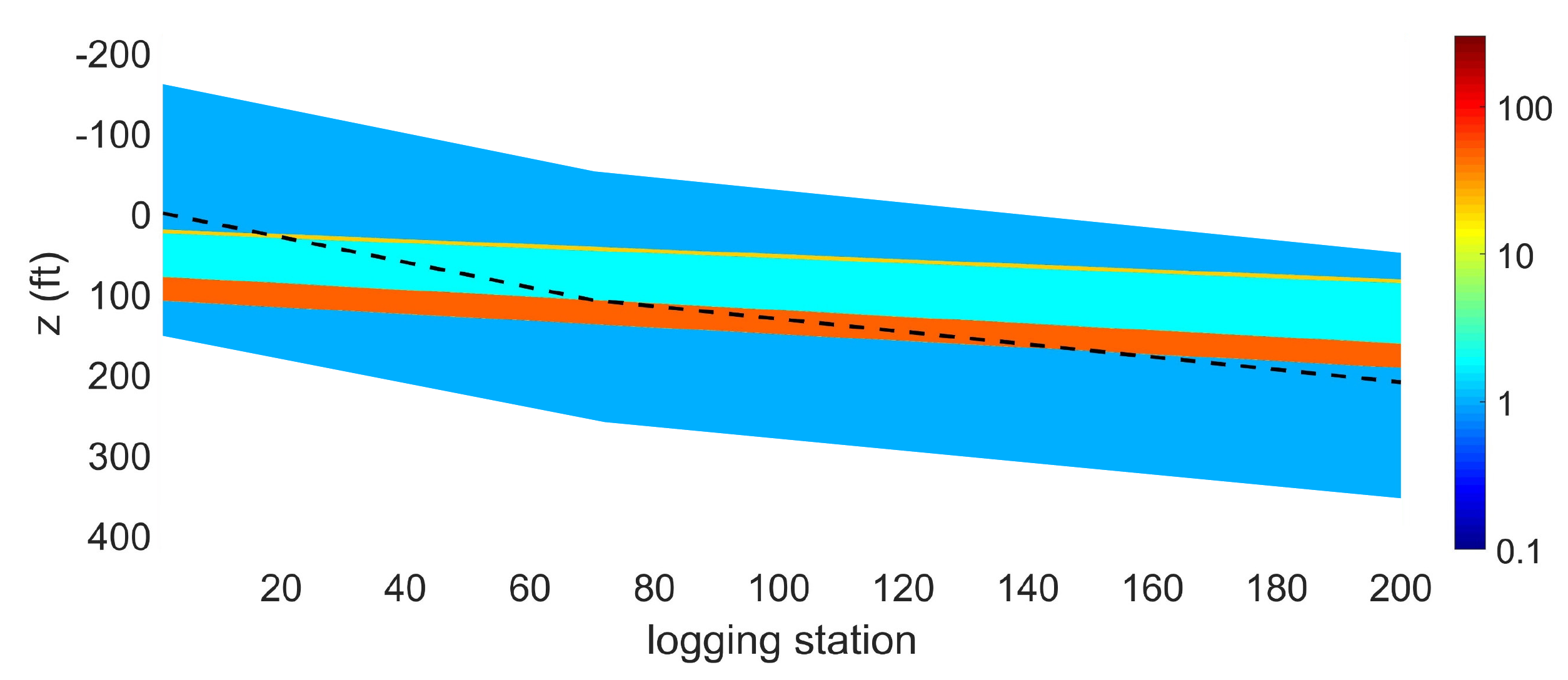}
\caption{A 5-layer earth model. Measurements are taken at 200 logging stations along the drilling trajectory represented by the black dot line.\label{fig:true_model}}
\end{figure}

We test a 5-layer earth model shown in Figure \ref{fig:true_model}. The ranges of model parameters are described in Table~\ref{table:true_model}. The DTB parameters in the table is calculated assuming that the tool is located at 0 ft. As the tool moves forward, the true DTB parameters for each logging station changes based on the tool location. The standard deviations of noises added to the measurements are also treated as unknown parameters, with the true values $\sigma_{ps}=0.75$ and $\sigma_{att}=0.125$ for the phase-shift and the attenuation signals, respectively.

\begin{table}[!ht]
\centering
\captionsetup{justification=centering}
\caption{True model parameters}
\begin{tabular}{cccccc}
\toprule
Layer index&1&2&3&4&5\\
\midrule
Rh ($\Omega\cdot m$) &  1  &  20  &  2  & 50 & 1     \\
DTB (ft) &  20-82  &  25-87  &  79-162  &  109-192 & - \\
thk (ft) &    -    &    5    &   54-75  &    30   & -\\
\bottomrule
\end{tabular}

\label{table:true_model}
\end{table}

\begin{table}[!ht]
\centering
\captionsetup{justification=centering}
\caption{Upper and lower boundaries of the uniform distribution}
\begin{tabular}{cccccc}
\toprule
Parameter & ln($\sigma_{ps}$) & ln($\sigma_{att}$) & ln(Rh) & $\rm{DTB}_1$ & thk\\
\midrule
min &  -10 & -10  &  -3 & -200 & 5   \\
max &  10  &  10  &  3  &  200 & 100 \\
\bottomrule
\end{tabular}
\label{table:prior}
\end{table}

We employ weakly informative uniform distributions for the model parameters, with bounds shown in Table~\ref{table:prior}. For the GTMCMC algorithm, the importance distribution for the first logging point is equal to the prior distribution, since there is no information from a previous logging point. Starting from the second logging point, the importance distribution is a multi-variate Gaussian distribution with the mean vector and covariance matrix computed from samples of the previous logging point. The coefficient of variance CoV is set to $0.2$ for all 1D inverse problems. We run the two algorithms on cluster with 200 Intel(R) Xeon(R) CPU E5-2650 v4 @2.20GHz CPUs with infiniBand interconnection, and $10^4$ samples are generated at each TMCMC stage so that each CPU only handles computations for $50$ samples. As a result, each inverse problem takes around 2 minutes to solve.

\begin{figure}[!ht]
    \centering
    \subfigure[]{
    \begin{minipage}[t]{0.49\linewidth}
    \centering
    \includegraphics[width=\columnwidth]{true_model.pdf}
    \end{minipage}
    }
    \subfigure[]{
    \begin{minipage}[t]{0.49\linewidth}
    \centering
    \includegraphics[width=\columnwidth]{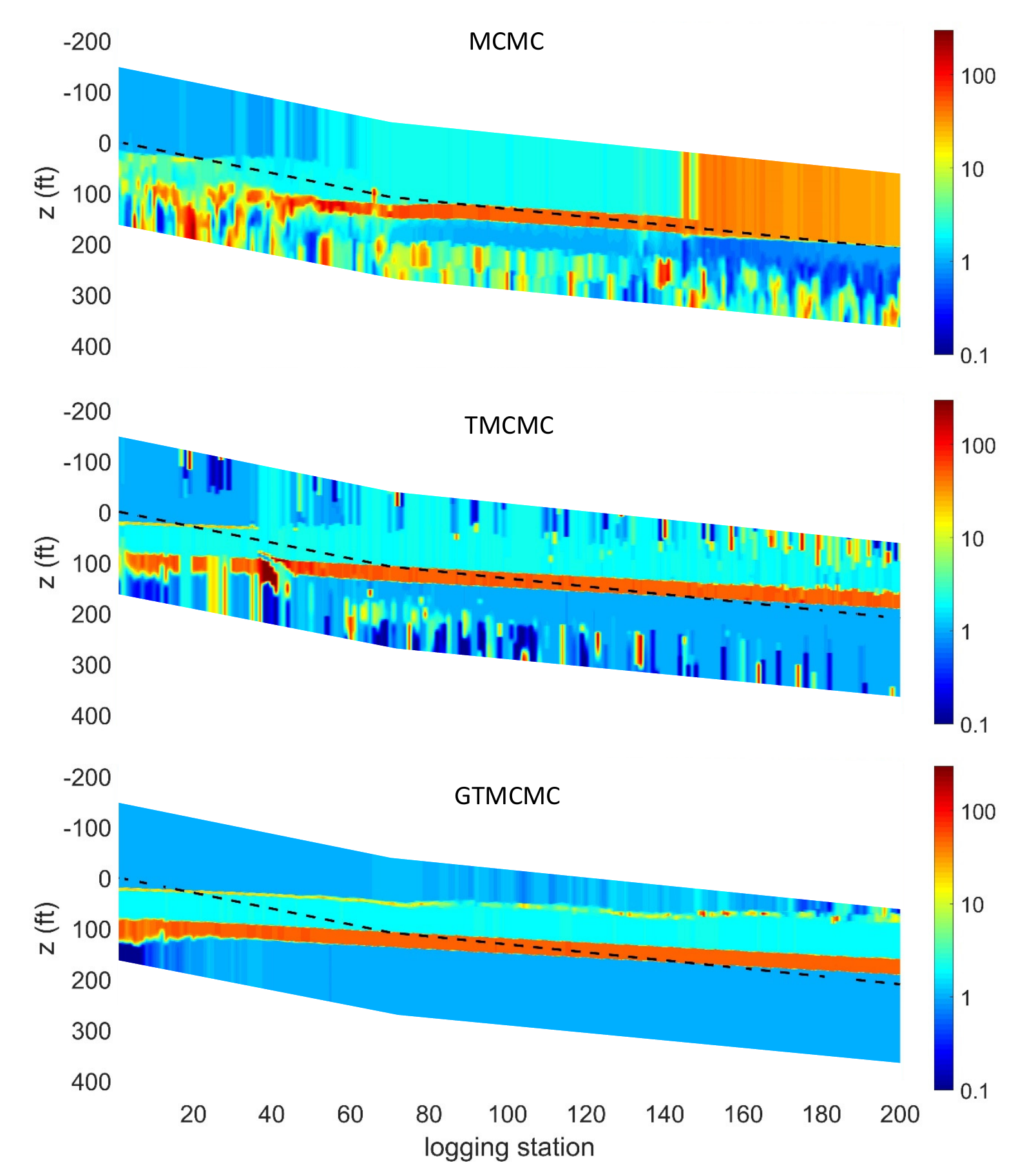}
    \end{minipage}
    }%
    \subfigure[]{
    \begin{minipage}[t]{0.49\linewidth}
    \centering
    \includegraphics[width=\columnwidth]{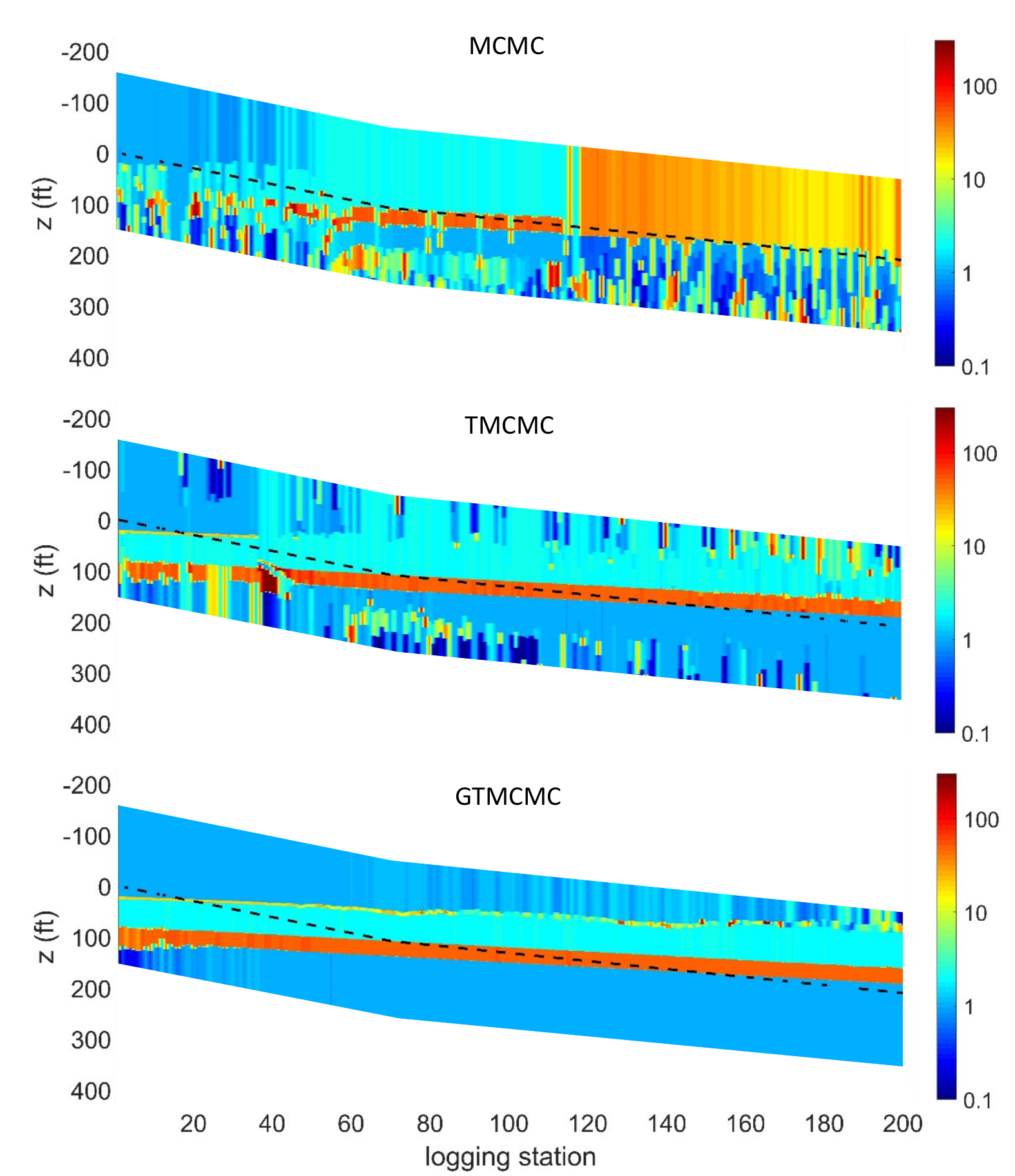}
    \end{minipage}
    }
    \caption{(a) The true model, (b) the mean prediction, and (c) the MAP prediction.}
    \label{fig:mean_and_MAP}
\end{figure}

In addition to sampling the posterior distribution with TMCMC and GTMCMC, we also run adaptive Metropolis MCMC as the baseline. 
For a fair comparison, we set the number of samples for each algorithm differently so that the total number of forward evaluations of solving $200$ problems are the same. As a result, the total number of forward evaluations are about $10^8$, which is $5\times10^5$ for each problem in average. Note that to get physically meaningful mean results, we discretize the models before computing the mean results. The mean and MAP results of the three algorithms are compared with the true model in Figure~\ref{fig:mean_and_MAP}. Given this sampling budget, the MCMC chains for high-dimensional LWD inverse problems are not converged. For all logging stations, only parameters of the closest layers can be inferred. The result from the TMCMC algorithm is more informative compared to the classical MCMC approach. With the same prior distribution at each logging station, the GTMCMC algorithm achieves better results than the TMCMC algorithm in the whole working range. Starting from logging station 40, the TMCMC algorithm starts to lose the identification of the top thin layer, which is only 5 ft thick. On the other hand, there exist sudden changes in the parameters between adjacent logging stations. For the GTMCMC algorithm, the initial samples are drawn from a importance distribution that is constructed with samples from the previous logging station, therefore, such sudden changes are limited. In addition, the GTMCMC algorithm can identify the top thin layer even though the logging tool is far from it. As a result, the 2D model resulted from the GTMCMC algorithm is smoother and visually more accurate than results generated by the MCMC and the TMCMC algorithms. 

\begin{figure}[!ht]
    \centering
     \subfigure[]{
    \begin{minipage}[t]{0.49\linewidth}
    \centering
    \includegraphics[width=\columnwidth]{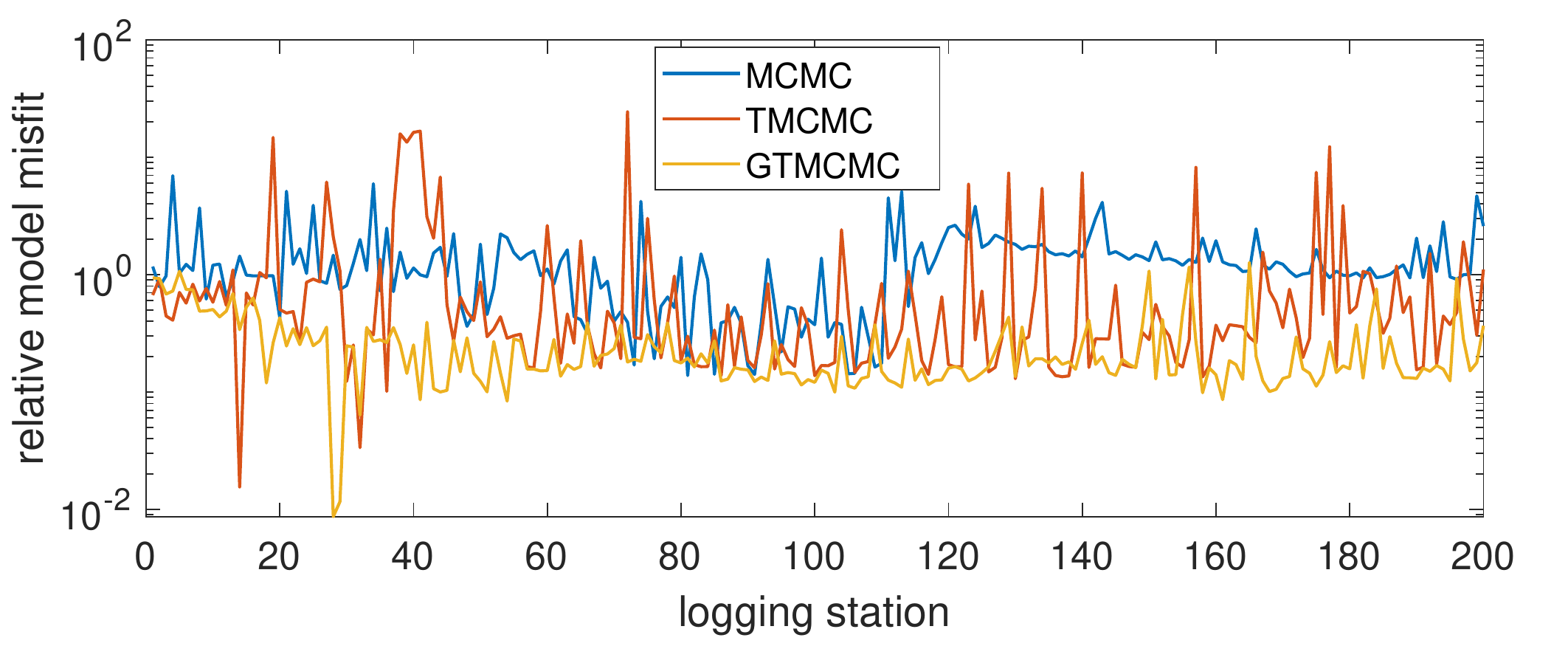}
    \end{minipage}
    }%
    \subfigure[]{
    \begin{minipage}[t]{0.49\linewidth}
    \centering
    \includegraphics[width=\columnwidth]{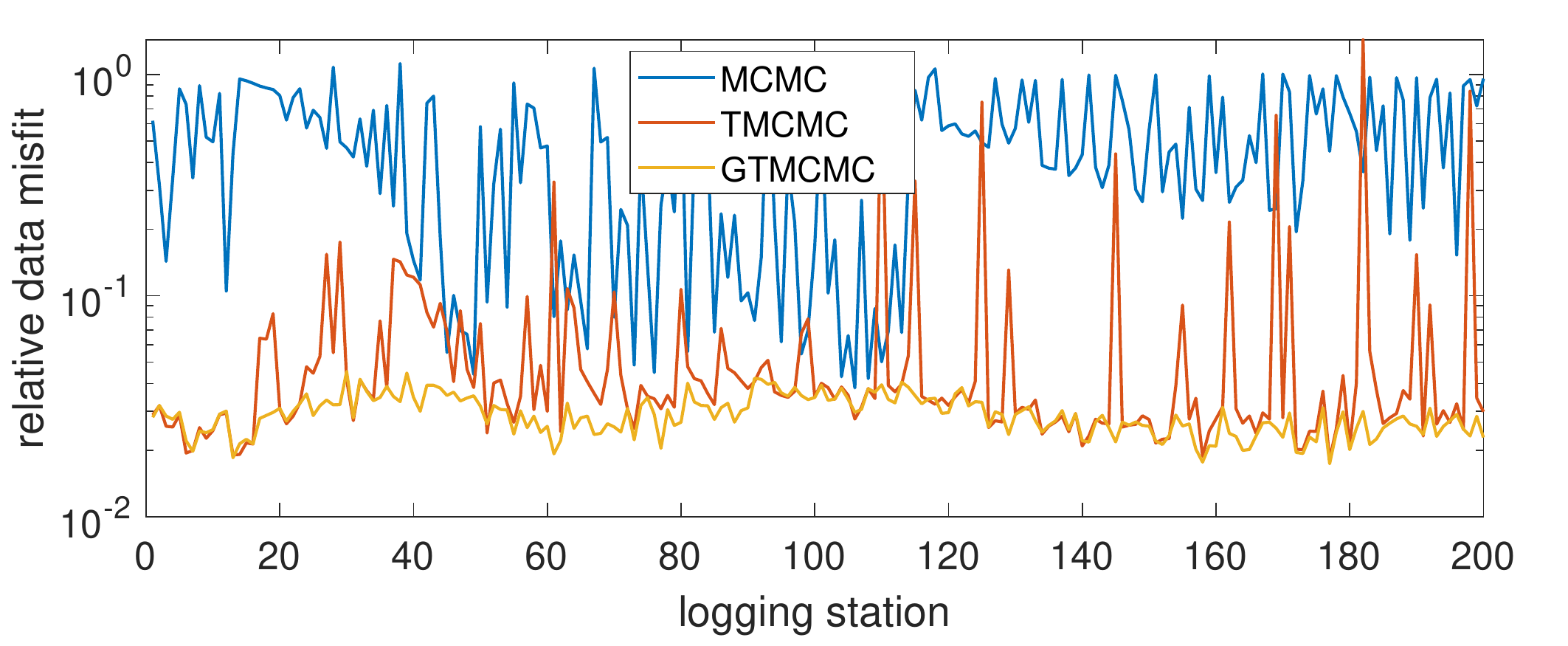}
    \end{minipage}
    }
    \caption{ Relative parameter (a) and data (b) misfits for the sampling algorithms tested in this section.}
    \label{fig:misfits}
\end{figure}
To evaluate the results obtained from the three sampling algorithms quantitatively, we calculate the parameter misfit and the model-data misfit of the mean models. In our experiments, the model parameters include resistivities and boundaries, so we convert the obtained models into pixel-based models to better calculate the parameter misfit. In the pixel-based model, the detection space is divided into 250 uniform thin layers so that the parameter only includes 250 resistivities. The relative parameter misfit and relative model-data misfit are calculated as follows

\begin{equation}
\centering
\text{relative parameter misfit} = \frac{\Vert \textbf{m}_{\text{true}} - \textbf{m}_{\text{inverse}}\Vert_2}{\Vert \textbf{m}_{\text{true}} \Vert_2}
\label{eq:model_misfit}
\end{equation}
\begin{equation}
\centering
\text{relative data misfit} = \frac{\Vert \textbf{d}_{\text{obs}} - \bm{f}(\textbf{m}_{\text{inverse}})\Vert_2}{\Vert \textbf{d}_{\text{obs}} \Vert_2}
\label{eq:data_misfit}
\end{equation}
with $\textbf{m}_{\text{true}}$ the true model parameters, $\textbf{m}_{\text{inverse}}$ the mean results from three sampling algorithms, $\textbf{d}_{\text{obs}}$ the noised observation and $\bm{f}(\textbf{m}_{\text{inverse}})$ the forward response of $\textbf{m}_{\text{inverse}}$.
We compute the relative model misfits and relative data misfits for 200 logging stations and shown them in Figure \ref{fig:misfits}.

\begin{figure}[!ht]
    \centering
     \subfigure[Posterior PDFs of resistivities]{
    \begin{minipage}[t]{0.6\linewidth}
    \centering
    \includegraphics[width=\columnwidth]{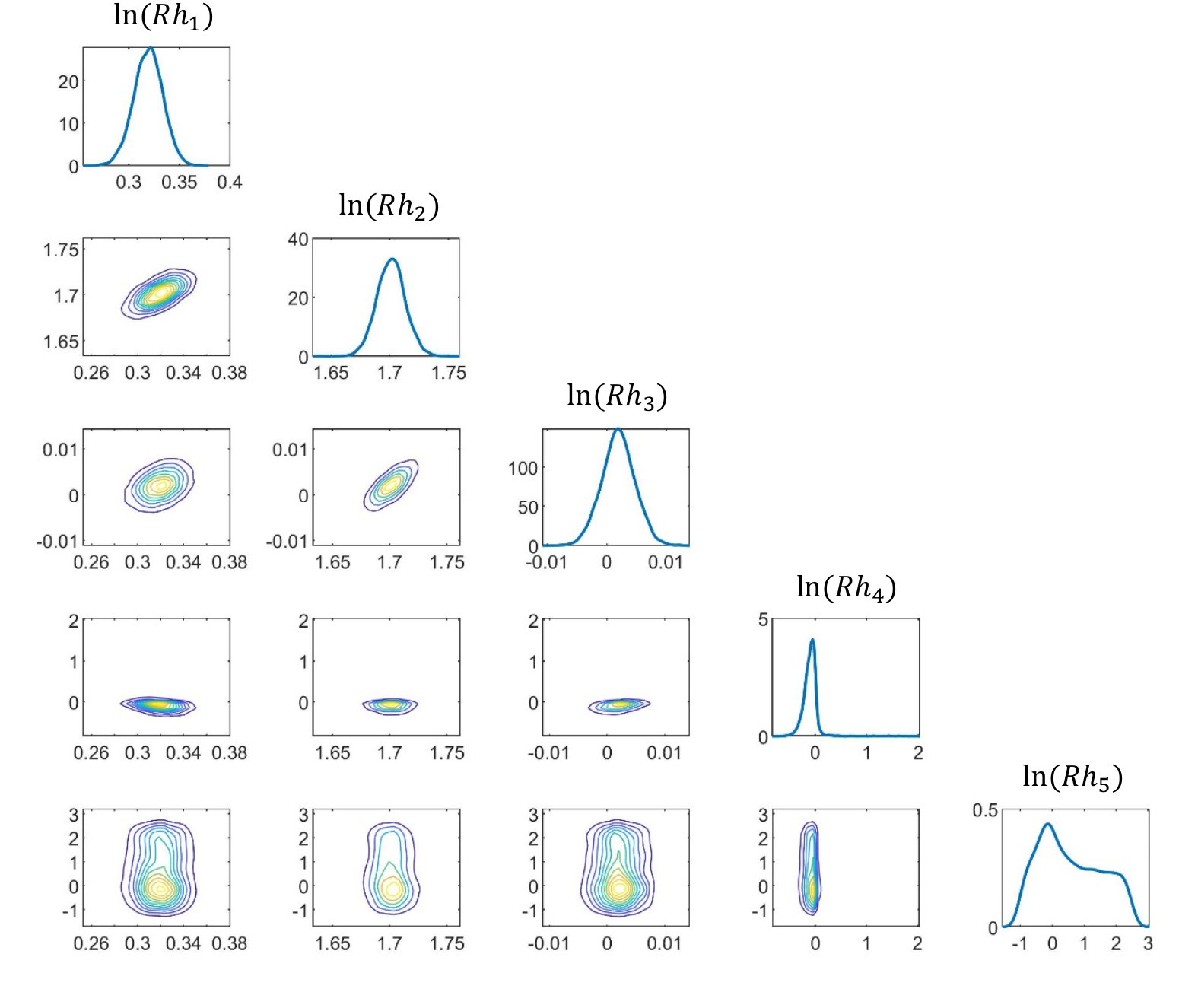}
    \end{minipage}
    }%
    \subfigure[Sample models]{
    \begin{minipage}[t]{0.3\linewidth}
    \centering
    \includegraphics[width=\columnwidth]{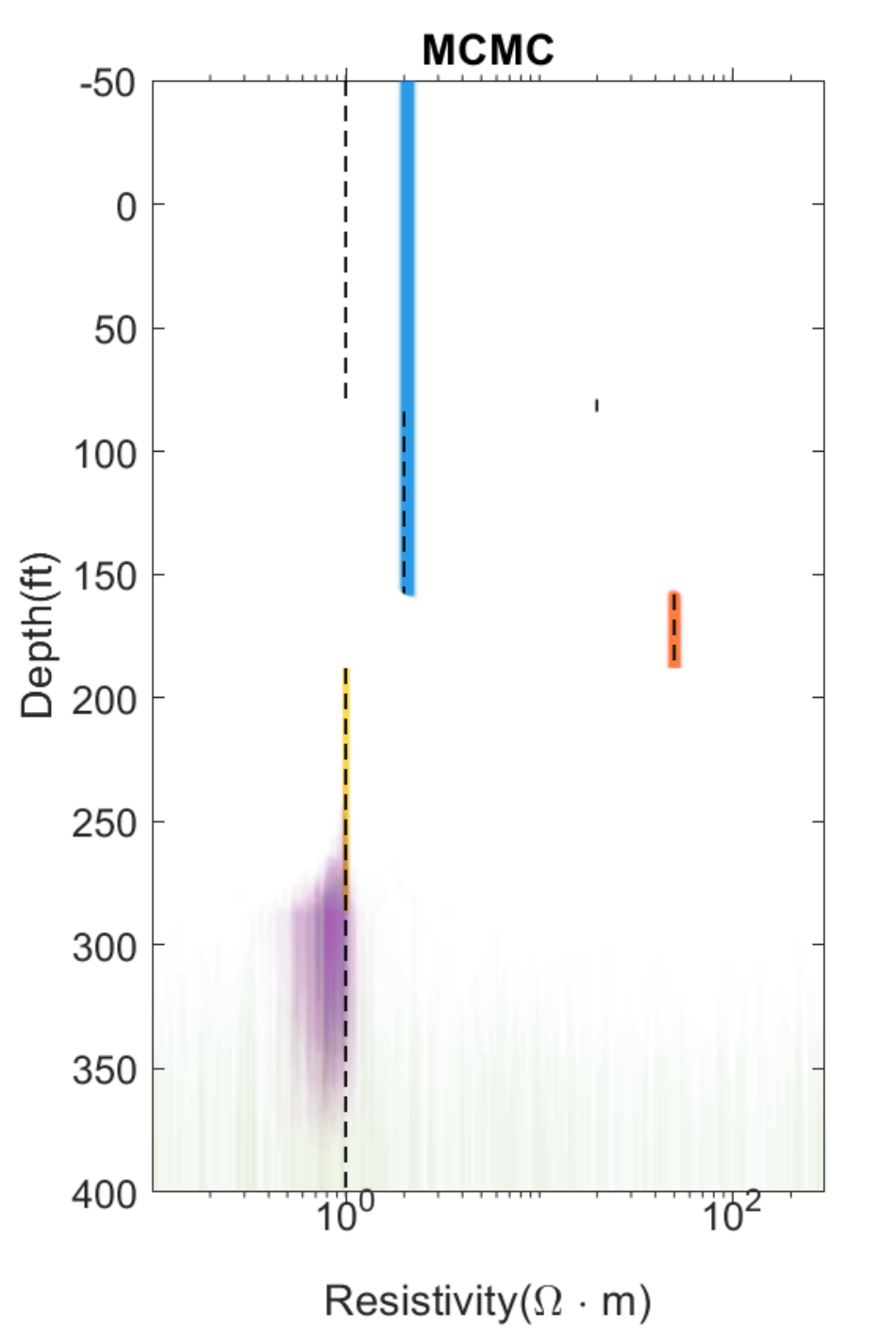}
    \end{minipage}
    }
    \caption{Statistics of the inversion result using MCMC at logging station 190. (a) Posterior PDFs of the resistivity parameters. (b) Image obtained by superimposing the values of resistivity in the layered media. The black dot line represents the ground truth.}
    \label{fig:P190MCMC}
\end{figure}

\begin{figure}[!ht]
    \centering
     \subfigure[Posterior PDFs of resistivities]{
    \begin{minipage}[t]{0.6\linewidth}
    \centering
    \includegraphics[width=\columnwidth]{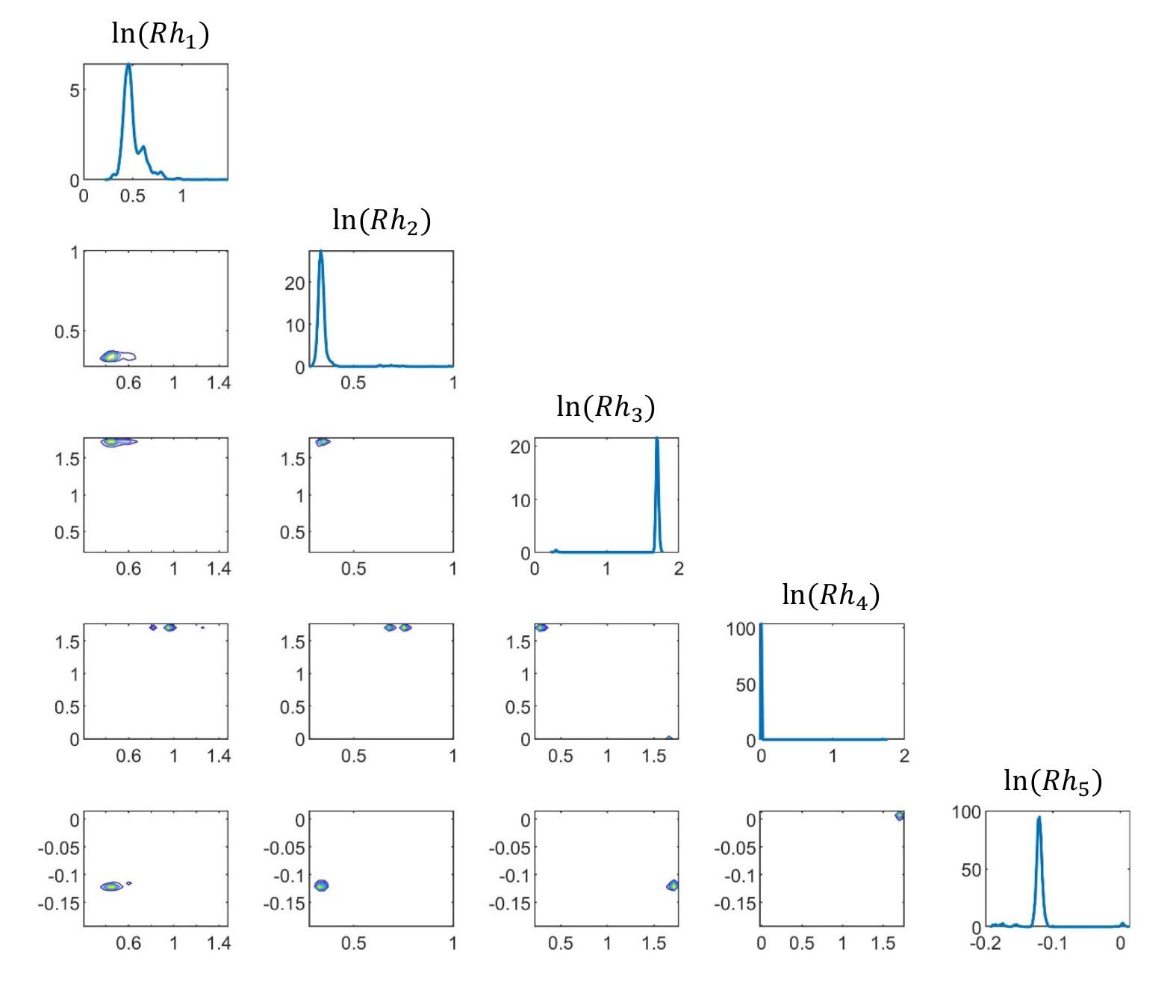}
    \end{minipage}
    }%
    \subfigure[Sample models]{
    \begin{minipage}[t]{0.3\linewidth}
    \centering
    \includegraphics[width=\columnwidth]{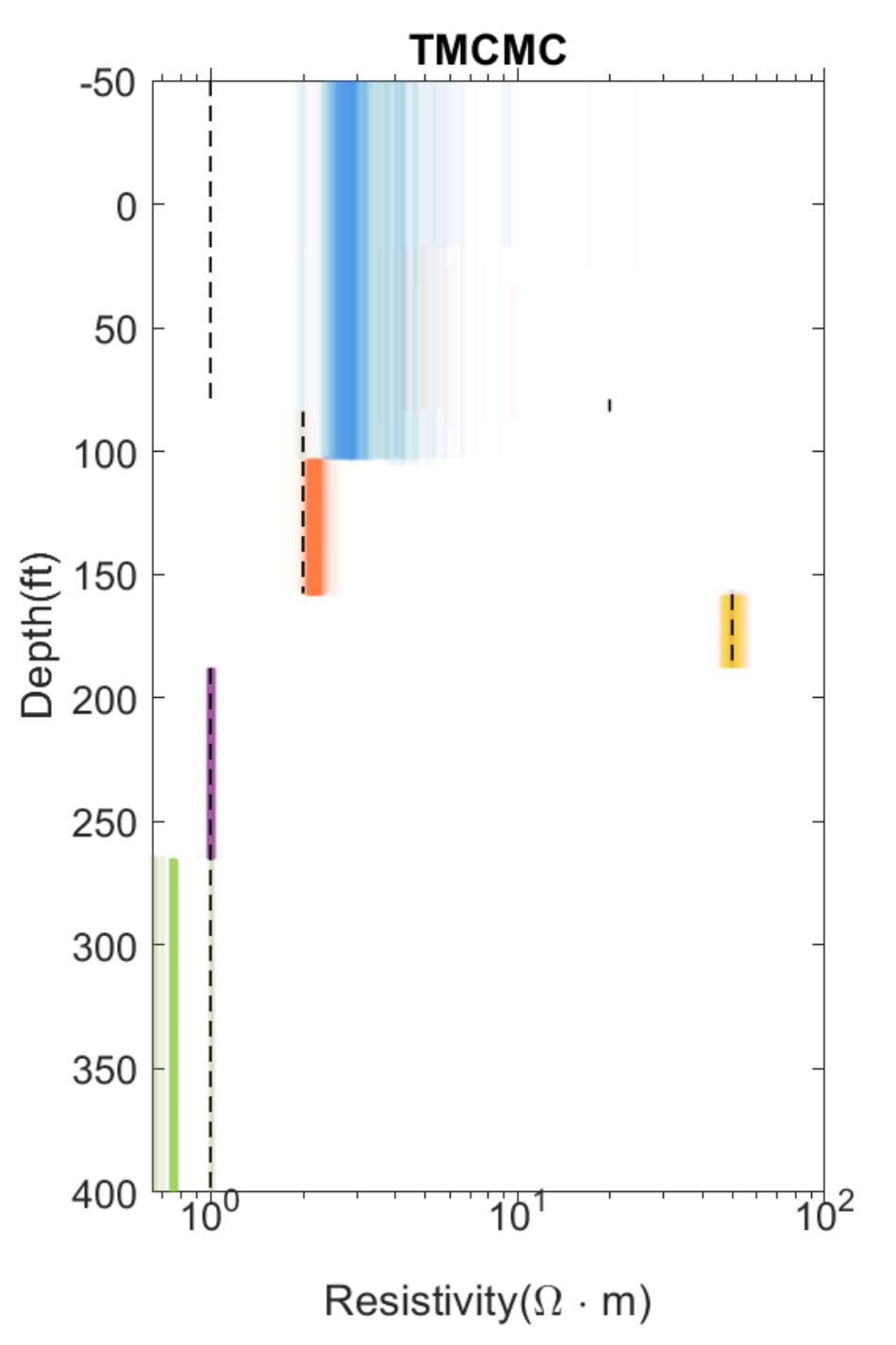}
    \end{minipage}
    }
    \caption{Statistics of the inversion result using TMCMC at logging station 190. (a) Posterior PDFs of the resistivity parameters. (b) Image obtained by superimposing the values of resistivity in the layered media. The black dot line represents the ground truth.}
    \label{fig:P190TMCMC}
\end{figure}

\begin{figure}[!ht]
    \centering
     \subfigure[Posterior PDFs of resistivities]{
    \begin{minipage}[t]{0.6\linewidth}
    \centering
    \includegraphics[width=\columnwidth]{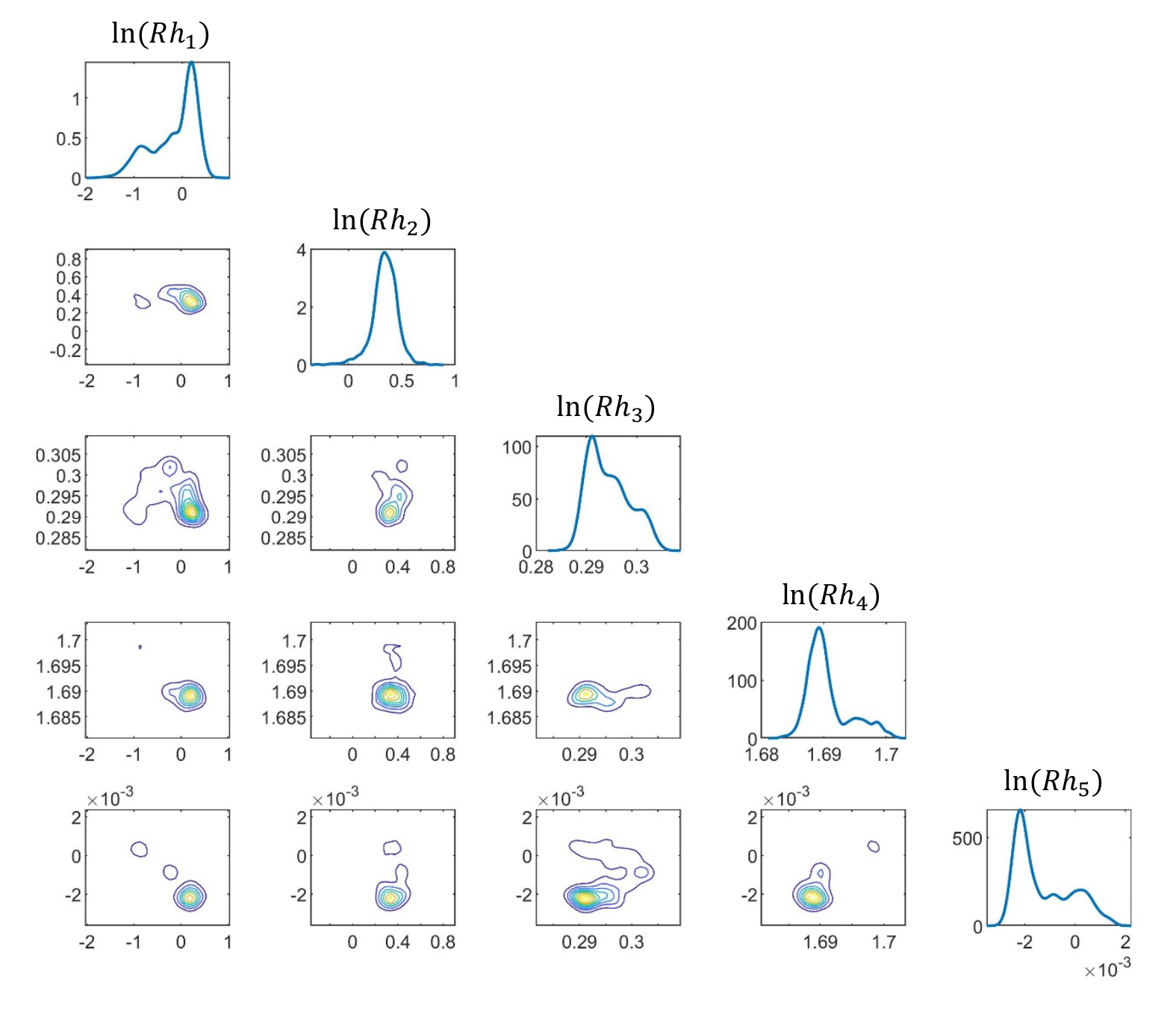}
    \end{minipage}
    }%
    \subfigure[Sample models]{
    \begin{minipage}[t]{0.3\linewidth}
    \centering
    \includegraphics[width=\columnwidth]{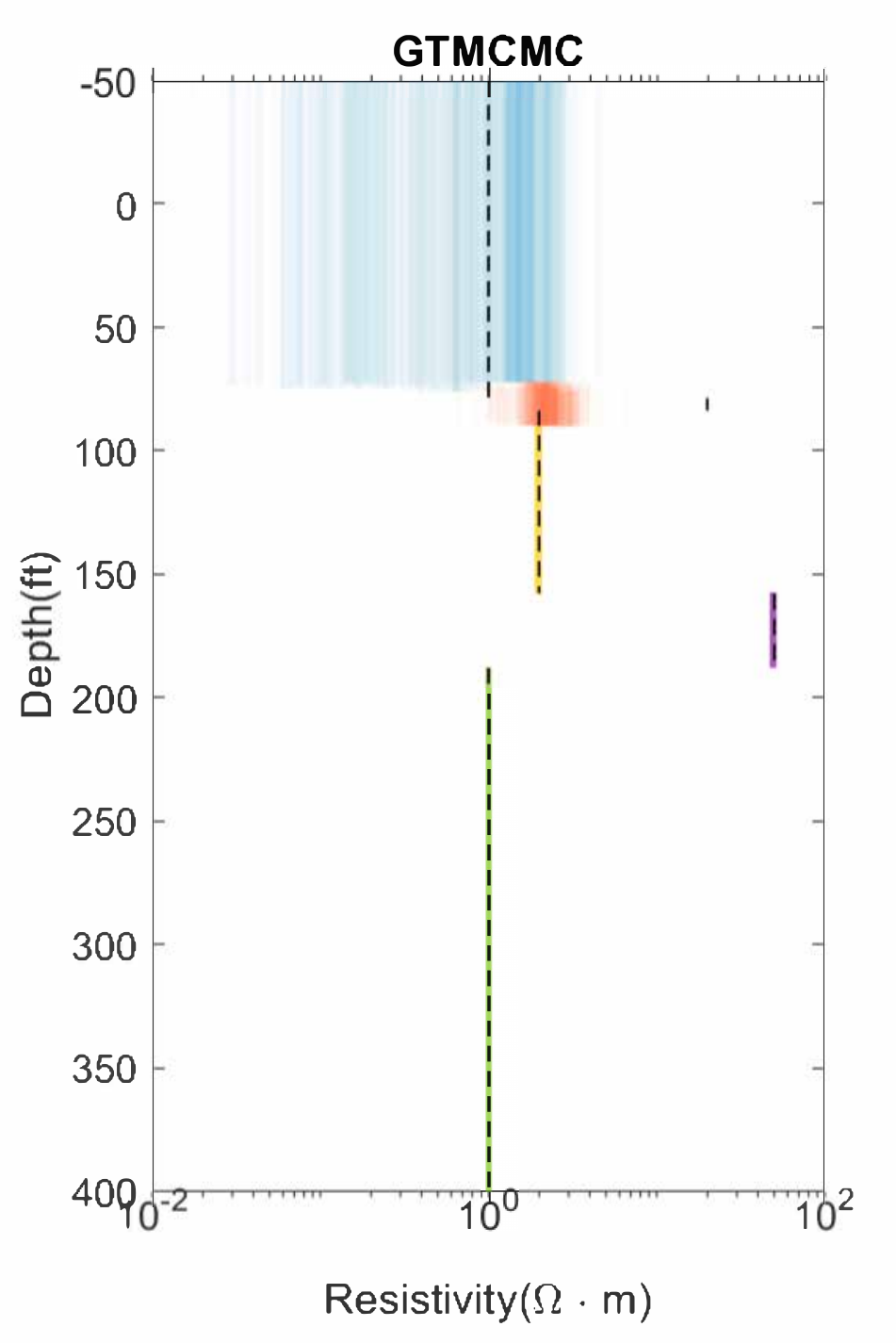}
    \end{minipage}
    }
    \caption{Statistics of the inversion result using GTMCMC at logging station 190. (a) Posterior PDFs of the resistivity parameters. (b) Image obtained by superimposing the values of resistivity in the layered media. The black dot line represents the ground truth. }
    \label{fig:P190PETMCMC}
\end{figure}

We examine results for logging station 190 as they are representative for the relative behavior of these models. Figures~\ref{fig:P190MCMC}(a), \ref{fig:P190TMCMC}(a), and ~\ref{fig:P190PETMCMC}(a) show the marginal posterior PDFs of the resistivity parameters in log scale. Figures~\ref{fig:P190MCMC}(b), \ref{fig:P190TMCMC}(b), and \ref{fig:P190PETMCMC}(b) show the image obtained by superimposing the values obtained for resistivity in the layered media when the tool is located at a depth of 200 ft. The vertical dot lines in the figures indicate the true resistivity of the layer. These results indicate different distributions of resistivities at logging station 190. From Figure~\ref{fig:P190MCMC}(b) we can recognize 3 layers, which correspond to the bottom 3 layers in the ground truth model. However, the boundaries of the first 2 layers are missed. Figure~\ref{fig:P190TMCMC}(b) shows that 5 layers can be recognized from the TMCMC samples, however, only the last 3 layers match the ground truth model. From Figure~\ref{fig:P190PETMCMC}(b) we recognize 5 layers, and the layer boundaries match the ground truth model well. The resistivities of the first two layers exhibit a higher uncertainty while the other three resistivities have very low uncertainty. These results indicate that the use of importance distributions from related logging stations can help increase accuracy for the inferred model parameters with a given computational budget.

\section{Conclusions}
\label{sec:conclusion}
Inspired by importance sampling, we presented a generalized transitional Markov chain Monte Carlo (GTMCMC) sampling strategy to improve the efficiency and generalize the applicability of TMCMC in tackling Bayesian inverse problems. With a judicious choice of importance PDF, GTMCMC is shown to reduce the otherwise prohibitive number of tempering stages when the prior PDF is significantly different from the target posterior. Furthermore, GTMCMC alleviates the challenge that TMCMC encounters when dealing with implicit priors as it no longer requires an initial set of samples from the prior PDF. Finally, the proposed generalization extends the applicability of TMCMC to inverse problems involving improper prior PDFs. Convergence analysis is provided in addition to numerical studies involving a variety of test problems, involving both Gaussian and non-Gaussian posterior PDFs, in which GTMCMC and TMCMC are compared and contrasted. The proposed GTMCMC algorithm was applied to a series of synthetic 1D logging-while-drilling inverse problems of relevance to the oil and gas industry, highlighting the enhanced efficiency and accuracy of GTMCMC over TMCMC and MCMC for such problems.

\section{Acknowledgements}
\label{sec:acknowledgements}
This material is based upon work supported by the U.S. Department of Energy, Office of Science, and Office of Advanced Science Computing Research, under Award Numbers DESC0017033. Sandia National Laboratories is a multimission laboratory managed and operated by National Technology and Engineering Solutions of Sandia, LLC, a wholly owned subsidiary of Honeywell International, Inc., for the U.S. Department of Energy’s National Nuclear Security Administration under contract DE-NA0003525. This paper describes objective technical results and analysis. Any subjective views or opinions that might be expressed in the paper do not necessarily represent the views of the U.S. Department of Energy or the United States Government.

\appendix

\section{Proof Monotonically Decreasing KL-Divergence}
\label{sec:info_mono}

\begin{lemma}
Let the intermediate PDFs $p_j \left (\theta \right)$ and $p_k \left (\theta \right)$ be defined as in Equation \eqref{eq:pjpdf} with annealing parameters $\beta_j$ and $\beta_k$ s.t. $\beta_j > \beta_k$ and $\beta_j, \beta_k \in \mathopen[0, 1]$. Further assume that $\text{D}_\text{KL} \left [p \left (\theta \mid D \right) \mid \mid p_0 \left (\theta \right) \right]$ is finite. Then $\text{D}_\text{KL} \left [p \left (\theta \mid D \right) \mid \mid p_j \left (\theta \right) \right] < \text{D}_\text{KL} \left [p \left (\theta \mid D \right) \mid \mid p_k \left (\theta \right) \right]$.
\label{lemma_kl}
\end{lemma}

\begin{proof}
Let us define an arbitrary intermediate distribution as:
\begin{equation}
p_{\beta} \left ( \theta \right )= \frac{\left (\frac{p \left (\theta \mid D \right)}{q \left (\theta \right)} \right )^{\beta} q \left (\theta \right)}{Z_{\beta}}.    
\end{equation}

Where the normalization is:
\begin{equation}
    Z_{\beta} = \int \left (\frac{p \left (\theta \mid D \right)}{q \left (\theta \right)} \right )^{\beta} q \left (\theta \right) d\theta.
\end{equation}

Then Lemma \ref{lemma_kl} $\Leftrightarrow \frac{\partial}{\partial \beta} \text{D}_\text{KL} \left [p \left (\theta \mid D \right) \mid \mid p_{\beta} \left (\theta \right) \right] < 0$, i.e. monotonically decreasing in $\beta$, $\forall \beta \in \mathopen[0,1]$.
\begin{align}
&\frac{\partial}{\partial \beta} \text{D}_\text{KL} \left [p \left (\theta \mid D \right) \mid \mid p_{\beta} \left (\theta \right) \right]  = \frac{\partial}{\partial \beta} \int p \left (\theta \mid D \right) \log \frac{p \left (\theta \mid D \right) }{p_{\beta} \left (\theta \right)} d\theta\\
&= - \int p \left (\theta \mid D \right) \log \frac{p \left (\theta \mid D \right) }{Q \left (\theta \right)} d\theta + \int p_{\beta} \left (\theta \right) \log \frac{p \left (\theta \mid D \right) }{Q \left (\theta \right)} d\theta\\
&= - \frac{1-\beta}{1-\beta} \left (\int p \left (\theta \mid D \right) \log \frac{p \left (\theta \mid D \right) }{Q \left (\theta \right)} d\theta - \int p_{\beta} \left (\theta \right) \log \frac{p \left (\theta \mid D \right) }{Q \left (\theta \right)} d\theta \right)\\
&=\frac{-1}{1-\beta} \left( \begin{array}{l}
	\int p \left (\theta \mid D \right) \log \frac{p \left (\theta \mid D \right) }{Q \left (\theta \right) \left (\frac{p \left (\theta \mid D \right) }{Q \left (\theta \right)} \right)^{\beta}} d\theta + \log{Z_{\beta}}  \\
	- \int p_{\beta} \left (\theta \right) \log \frac{p \left (\theta \mid D \right) }{Q \left (\theta \right) \left (\frac{p \left (\theta \mid D \right) }{Q \left (\theta \right)} \right)^{\beta}} d\theta - \log{Z_{\beta}}
\end{array}  \right)\\
&=\frac{-1}{1-\beta} \left (\int p \left (\theta \mid D \right) \log \frac{p \left (\theta \mid D \right) }{p_{\beta} \left (\theta \right)} d\theta - \int p_{\beta} \left (\theta \right) \log \frac{p \left (\theta \mid D \right) }{p_{\beta} \left (\theta \right)} d\theta \right )\\
&=\frac{-1}{1-\beta} \left ( \text{D}_\text{KL} \left [p \left (\theta \mid D \right) \mid \mid p_{\beta} \left (\theta \right) \right] + \text{D}_\text{KL} \left [ p_{\beta} \left (\theta \right) \mid \mid p \left (\theta \mid D \right) \right]\right)\\
& \leq 0
\end{align}

This follows because the KL divergence is always non-negative and $1-\beta$ is always non-negative for $\beta \in \mathopen[0,1]$. This holds with equality only when $\text{D}_\text{KL} \left [ p_{\beta} \left (\theta \right) \mid \mid p \left (\theta \mid D \right) \right] = 0$ and $\text{D}_\text{KL} \left [ p_{\beta} \left (\theta \right) \mid \mid p \left (\theta \mid D \right) \right] = 0$ which occurs when $\beta =1$ and $\text{D}_\text{KL} \left [p \left (\theta \mid D \right) \mid \mid p_0 \left (\theta \right) \right]$ is finite.

\end{proof}

Lemma \ref{lemma_kl} means that the intermediate PDF $p_j \left (\theta \right)$ is closer to the posterior, $p \left (\theta \mid D \right)$, than $p_k \left (\theta \right)$ in terms of the informational theoretic Kullback-Leibler divergence. Since this holds for any pair $\beta_j$ and $\beta_k$ s.t. $\beta_j > \beta_k$ and $\beta_j, \beta_k \in \mathopen[0, 1]$, then we can conclude information is always being gained about the posterior through annealing. Of course for practical purposes, this assumes that the samples is adequately capturing the intermediate distribution which is asymptotically true as the number of samples grows.

\section{Proof Monotonically Increasing CoV}
\label{sec:cov_mono}

\begin{lemma}
Let the importance distribution $Q \left (\theta \right)$ be the initial distribution and the posterior $p \left (\theta \mid D \right)$ be the target distribution. Assume that $\text{D}_\text{KL} \left [p \left (\theta \mid D \right) \mid \mid Q \left (\theta \right) \right]$ is finite. Then $\forall \beta \in \mathopen(0,1)$ the amount of information gained in the update from the initial distribution to the target distribution in the view of the intermediate distribution $p_{\beta} \left (\theta \right)$ is monotonically increasing in $\beta$.
\label{lemma_info2}
\end{lemma}

\begin{proof}
The information gain from the initial distribution, $Q \left (\theta \right)$, to the target, $p \left (\theta \mid D \right)$ in the view of $p_{\beta} \left (\theta \right)$ is defined as:
\begin{equation}
\mathcal{I}_{p_{\beta} \left (\theta \right)} \left [p \left (\theta \mid D \right) \mid \mid Q \left (\theta \right)\right ] = \int p_{\beta} \left (\theta \right) \log \frac{p \left (\theta \mid D \right)}{Q \left (\theta \right)} d\theta
\end{equation}

Therefore,
\begin{align}
\frac{\partial}{\partial \beta}\mathcal{I}_{p_{\beta} \left (\theta \right)} \left [p \left (\theta \mid D \right) \mid \mid Q \left (\theta \right)\right ] &= \frac{\partial}{\partial \beta} \int p_{\beta} \left (\theta \right) \log \frac{p \left (\theta \mid D \right)}{Q \left (\theta \right)} d\theta\\ \nonumber
&=\int p_{\beta} \left (\theta \right) \left ( \log \frac{p \left (\theta \mid D \right)}{Q \left (\theta \right)} \right)^2 d\theta \\
& \ \ \  - \left (\int p_{\beta} \left (\theta \right) \log \frac{p \left (\theta \mid D \right)}{Q \left (\theta \right)} d\theta \right )^2\\
&\geq 0
\end{align}

This inequality follows from Jensen's inequality. Thus $\mathcal{I}_{p_{\beta} \left (\theta \right)} \left [p \left (\theta \mid D \right) \mid \mid Q \left (\theta \right)\right ]$ is monotonically increasing. Note that this line of reasoning holds also for a discrete random variable and where the integral is replaced by a sum.
\end{proof}

\begin{lemma}
The coefficient of variation, $\kappa$, of sample weights defined by a change in $\beta$ is monotonically increasing as $\beta$ increases.
\end{lemma}

\begin{proof}
The coefficient of variation is defined as:
\begin{equation}
\kappa = \frac{\sqrt{\int \left (\frac{p \left (\theta \mid D \right)}{Q \left (\theta \right)}\right )^{2\beta} Q \left (\theta \right)d\theta -\left (\int \left (\frac{p \left (\theta \mid D \right)}{Q \left (\theta \right)}\right )^{\beta} Q \left (\theta \right ) d\theta \right )^2}}{\int \left (\frac{p \left (\theta \mid D \right)}{Q \left (\theta \right)}\right )^{\beta} Q \left (\theta \right) d\theta}
\end{equation}

Therefore,
\begin{equation}
1+\kappa^2 = \frac{\int \left (\frac{p \left (\theta \mid D \right)}{Q \left (\theta \right)}\right )^{2\beta} Q \left (\theta \right)d\theta}{\left (\int \left (\frac{p \left (\theta \mid D \right)}{Q \left (\theta \right)}\right )^{\beta} Q \left (\theta \right) d\theta \right )^2}
\end{equation}

Differentiating with respect to $\beta$, we find:
\begin{align}
\frac{\partial \kappa}{\partial \beta} &= \frac{1+\kappa^2}{\kappa} \left( 
\begin{array}{l}
	\frac{\int \left (\frac{p \left (\theta \mid D \right)}{Q \left (\theta \right)}\right )^{2\beta} Q \left (\theta \right) \log \frac{p \left (\theta \mid D \right)}{Q \left (\theta \right)} d\theta}{\int \left (\frac{p \left (\theta \mid D \right)}{Q \left (\theta \right)}\right )^{2\beta} Q \left (\theta \right)d\theta
	} \\
- \frac{\int \left (\frac{p \left (\theta \mid D \right)}{Q \left (\theta \right)}\right )^{\beta} Q \left (\theta \right) \log \frac{p \left (\theta \mid D \right)}{Q \left (\theta \right)} d\theta}{\int \left (\frac{p \left (\theta \mid D \right)}{Q \left (\theta \right)}\right )^{\beta} Q \left (\theta \right) d\theta}
\end{array} \right )\\
&= \frac{1+\kappa^2}{\kappa} \left (\int p_{2\beta} \left (\theta \right) \log \frac{p \left (\theta \mid D \right)}{Q \left (\theta \right)} d\theta - \int p_{\beta} \left (\theta \right) \log \frac{p \left (\theta \mid D \right)}{Q \left (\theta \right)} d\theta \right )\\
& \geq 0
\end{align}

This follows because $\kappa$ is always positive and by the monotonic increase in information described by Lemma \ref{lemma_info2}. Note that again we can follow a similar line of reasoning when considering the discrete estimation of the coefficient of variation where the mean and variance of the weights are estimated using samples and therefore the integrals are sums. Further, since the choice of $Q \left (\theta \right)$ is arbitrary, this holds of all stages of the algorithm.
\end{proof}

Because $\kappa$ is monotonic then it can be easily optimized to a target level of variation using approaches like a bisector method.

\normalem



\bibliographystyle{unsrt}

\end{document}